The brittle-plastic transition in quartz-albite mixtures: New insights from shear deformation experiments at mid-to-lower crustal depth conditions

Miho Furukawa[1], Berend A. Verberne[2], Sando Sawa[1], Hiroyuki Nagahama[1], Miki Takahashi[2], Oliver Plümper[3,4], Jun Muto[1]

[1] Department of Earth Science, Graduate School of Science, Tohoku University, Sendai, Japan.

[2] Geological Survey of Japan, The National Institute of Advanced Industrial Science and Technology, Tsukuba, Japan.

[3] Department of Earth Sciences, Utrecht University, Utrecht, The Netherlands.

[4] Faculty of Geosciences and MARUM - Center for Marine Environmental Sciences, University of Bremen, Germany.

**Key Points:**

- Shear experiments at various simulated depth conditions provide microphysical evidence for brittle-plastic transition in the upper crust.
- Grain boundary sliding in nano-grain domains controls crustal strength in the brittle-plastic transition zone.
- Grain boundary sliding and dynamic recrystallization reduce crustal strength at the base of the seismogenic zone.

**Abstract**

Crustal strength is often characterized using a strength-depth profile where laboratory-derived friction and flow laws are connected at depth (i.e., the so-called "Christmas Tree" diagram). Large, destructive earthquakes frequently nucleate within the transition zone from a frictional-to-viscous deformation regime, which represents the strongest part of the crust. However, microscale deformation mechanisms controlling bulk frictional-to-viscous transitional behavior remain unclear. To investigate the deformation mechanisms, we conducted shear experiments on room-dry, powdered quartz-albite mixtures under upper- to mid-crustal pressure-temperature conditions using a Griggs-type deformation apparatus. We simulated depth conditions in the range 7-30 km, by varying temperatures and confining pressures (210-900 °C and 185-870 MPa, respectively, by assuming 30 °C/km and 2700 kg/m$^3$). To assess the rate dependence and stability of shear deformation, we sequentially stepped shear strain rates between ~10$^{-3}$ /s and ~10$^{-4}$ /s. At shallower depth conditions, friction coefficients follow Byerlee's law, while at greater depth conditions they deviate from it and strain weakening is observed. Post-mortem microstructures indicate changing deformation mechanisms with increasing simulated depths. The samples deformed at shallower depth conditions (<18 km) show a predominance of cataclastic grain comminution. At greater depth conditions (>24 km), nano-grains are observed, as well as polygonal quartz grains at the greatest depth condition (30 km). These results indicate that the controlling deformation mechanisms at the frictional-viscous transition zone are grain boundary sliding and dynamic recrystallization. We conclude that nano-scale deformation mechanisms govern the frictional-viscous transitional deformation in the upper crust, and propose their importance for understanding seismic rupture processes there.

**Plain Language Summary**

Laboratory studies show that rocks in the Earth's crust change how they deform with depth. Near the surface, they tend to break apart (brittle behavior), while deeper down they slowly flow (plastic behavior). Large inland earthquakes often begin in the transition zone between these two styles, but it is unclear how the two interact there. To investigate this subject, we recreated fault zones in the lab using mixtures of quartz and albite, and deformed them under different pressures and temperatures that represent increasing depth. At shallow conditions, the samples became stronger as they were deformed, with grains breaking into small fragments. At greater depths, the samples instead weakened with deformation, and their grains stretched and even shrank to nanometer sizes. These results show that the brittle-to-plastic change happens gradually through the transition zone, and that extreme grain size reduction plays a key role in weakening the crust and in the processes that lead to earthquake nucleation.

## 1. Introduction

Crustal strength in the upper crust has been characterized by empirically derived friction and flow laws of fault rocks. At shallow depths, fault strength is constrained using Byerlee's law (Byerlee, 1978), whereas at greater depths, a plastic flow law (e.g., Goetze, 1971) is used. Connecting the frictional and flow laws yields a crustal strength profile, which is frequently referred to as the "Christmas Tree diagram" (e.g., Brace & Kohlstedt, 1980; Scholz, 1988; Kohlstedt et al., 1995; Rutter et al., 2001; Ellis & Wang, 2022). This model has proven extremely useful in better determining crustal deformation behavior, through, e.g., lithosphere dynamics (e.g., Bird, 1978; Bürgmann & Dresen, 2008; Shibazaki et al., 2008, 2016; Muto, 2011) and earthquake cycle modeling (e.g., Ellis & Stöckhert, 2004; Shimamoto & Noda, 2014; Muto et al., 2016; Van den Ende et al., 2020). However, in its simplest form, the Christmas Tree inherently overestimates fault strength at the projected intersection between frictional fault slip and plastic flow (Shimamoto, 1986), that is, at the brittle-plastic or frictional-viscous transition.

Transitional brittle-ductile deformation behavior has been thoroughly investigated in simulated fault slip experiments using granular halite aggregates (Shimamoto, 1986, 1989; Hiraga & Shimamoto, 1987; Chester, 1988; Kawamoto, 1996; Bos & Spiers, 2001; Niemeijer & Spiers, 2006; Takahashi et al., 2017; Hirauchi et al., 2020). The experiments using halite illuminated the presence of wide "semi-brittle" and "semi-ductile" deformation regimes in between "pure brittle" or "frictional" and "pure ductile" or "plastic" regimes (Shimamoto, 1986, 1989). Specifically, the peak strength achieved in the brittle-ductile transition is much lower than that projected from the simple Christmas Tree. To capture this transitional behavior quantitatively, Noda & Shimamoto (2012) proposed an empirical law by smoothly connecting the rate- and state-dependent friction law (Dieterich, 1979; Ruina, 1983) with the power-law flow law by Noda & Shimamoto (2010). Deformation experiments at hydrothermal conditions on more representative crustal fault rocks such as granitic (Blanpied et al., 1991, 1995) and clayey (Den Hartog et al., 2012a, b, 2013; Den Hartog & Spiers, 2013) compositions also show mechanical behavior reminiscent of a frictional-to-viscous transition. For fault gouges composed of calcite, Verberne et al. (2014a, b, 2017) reported localized nanogranular slip zones characterized by flow structures suggestive of ductile deformation, even at room temperature conditions. This observation suggests an essential role of grain size (evolution) in controlling fault slip and seismogenesis (Verberne et al., 2014a, 2019, 2020). A model based on a competition between time-sensitive versus time-insensitive intergranular deformation processes (Bos & Spiers, 2001; Niemeijer & Spiers, 2006, 2007; Chen & Spiers, 2016) can reproduce the shear behavior of calcite gouge (Chen et al., 2020), and successfully capture the microphysical processes of fault evolution with slip rate (Chen et al., 2021, 2022; Mei & Rudnicki, 2023).

High-pressure and high-temperature deformation experiments, aimed at investigating deformation mechanisms under mid- to lower crustal conditions, have been frequently carried out using Griggs-type, solid confining medium loading apparatuses (e.g., Kronenberg & Shelton, 1980; Hirth & Tullis, 1992; Muto et al., 2011; Peč et al. 2016; Précigout et al., 2019; Okazaki & Hirth, 2020). Tullis & Yund (1977) performed a series of deformation experiments of granitic cylinders under conditions covering the brittle-ductile transition, revealing a systematic change in microstructures with increasing temperature and confining pressure. Tullis & Yund (1987) further investigated deformation mechanisms in albite at the lower temperature region of the ductile regime. Experiments on the brittle-ductile transition of anorthosite (Hadizadeh & Tullis, 1992; Tullis & Yund, 1992) showed similar behaviors to those in albite. Meanwhile, Hirth & Tullis (1989) found that at room temperature, a transition from faulting to cataclastic flow of quartzite occurs at relatively lower pressure in the presence of sufficient pores. At elevated temperatures, Hirth & Tullis (1992) proposed that quartz aggregates undergo distinctive dislocation creep regimes depending on temperature and strain rate. Tullis et al. (1990) investigated the effect of pre-existing faults on forming localized ductile shear zones in feldspathic rocks. Furthermore, Dell'Angelo & Tullis (1996) concluded that strain weakening of quartz-feldspar aggregates are governed by grain boundary migration of feldspars, which is combined with interconnection of the weaker quartz grains at higher temperature and higher strains. More recently, Griggs-type experiments on granitoid gouges demonstrated the formation of nanocrystalline and amorphous layers, which may play a crucial role in controlling weakening toward seismic failure (Peč et al., 2016). Similarly, amorphous materials were pointed out to be a potential trigger of earthquakes in experiments of quartz (Nakamura et al., 2012) and lawsonite (Shiraishi et al., 2022). On the other hand, weakening at much greater depths —where plastic deformation dominates, and seismic failure is suppressed — is controlled by dynamic recrystallization (Hirth & Tullis, 1992). Summarizing these studies indicate that (fault) rock-weakening mechanisms are depth dependent: at shallower depths, formation of nanocrystalline and amorphous material provides weak domains resulting in shear instabilities potentially leading to fault slip, while at greater depths, dynamic recrystallization dominates and viscous creep persists rather than fracturing. However, microscale deformation mechanisms that govern bulk frictional-viscous transitional behavior remains poorly understood, since experimental studies under a wide range of temperature and confining pressure are limited for the crust-composing quartz-feldspar mixtures. To reveal such microscale transition, deformation experiments under various conditions that span the entire frictional-viscous transitional condition are required.

Aiming at revealing microscale deformation mechanisms at the brittle-plastic transition zone in the continental upper crust, we performed general shear deformation experiments with focus on a microstructural change with depth, from gouges and cataclasites at brittle deformation conditions to mylonites at plastic deformation conditions. Using a Griggs-type deformation apparatus, we deformed quartz-albite mixtures under various simulated depth conditions covering the brittle-plastic transition zone. Quartz and feldspar represent a granodioritic composition, which is the most abundant mineral composition in Earth's crust (Rudnick & Fountain, 1995). Mechanical data from experiments is integrated with post-mortem microstructural analysis using transmission and scanning electron microscopy. Our results indicate that the dominant deformation mechanism gradually changes with depth, from frictional slip at shallower depths to grain boundary sliding between nano-grains and dynamic recrystallization at greater depths. We conclude that nanogranular slip zones play an important role in controlling crustal strength at the brittle-plastic transition zone in the upper crust.

## 2. Materials and methods

## 2.1. Sample materials

The samples in our experiments are composed of mixtures of quartz (JIS class 2, the association of powder process industry and engineering, Japan) and albite (FN100, Kyoritsu material, Japan, the same source as Shigematsu et al., 2022) powders, both of which are commercially available. To remove organic impurities, albite FN100 was dried in an oven for about two hours at 600 °C (so-called dewaxing, Shigematsu et al., 2022). Small amounts of apatite, K-feldspar, and Mg-containing particles are included, which appear bright on a backscattered electron (BSE) image. Their cumulative area on a BSE image is fewer than 5 % (Fig. 1d); therefore, these impurities are deemed insignificant to the mechanical behavior. However, they are useful as shear sense indicators using the direction of elongation. Laser particle size analysis showed that a geometric mean characterizes the grain size distribution of the quartz powder ($\overline{d_{\mathrm{G}}}$) of 26.3 μm, geometric standard deviation ($\mathrm{SD_G}$) of 2.5 μm, and median ($d_{50}$) of 27.2 μm. The grain size distribution of the albite powder is characterized by $\overline{d_{\mathrm{G}}}$ = 37.7 μm, $\mathrm{SD_G}$ = 2.5 μm, and $d_{50}$ = 46.3 μm. (Fig. S1). The powders were mixed at a mass ratio of quartz: albite = 1: 1. Each experiment used ~ 0.1 g of the powder mixture.

## 2.2. Deformation experiments

### 2.2.1. Experimental conditions

Deformation experiments were conducted using a Griggs-type solid medium apparatus installed at Tohoku University (Fig. 1). We used a solid salt assembly (SSA in Holyoke & Kronenberg, 2010) in all the experiments (see Fig. 1d for details). Kido et al. (2016) reported that the stresses measured with the SSA are consistent with those obtained in Holyoke & Kronenberg (2010) within ± 30 MPa under identical conditions. The sample was placed evenly between a pair of alumina pistons cut at 45° with respect to the direction of the loading axis. The pre-cut surfaces were manually grooved using a diamond blade, to achieve a firm grip between the sample powder and pistons. The sample and pistons were encapsulated in an annealed Ag or Pt jacket. Both ends of the jacket were bent and mechanically sealed with Ag or Pt disks. Detailed experimental methods are explained in Text S1. To simulate realistic crustal conditions, we varied temperature ($T$) and confining pressure ($P_c$) by assuming a geothermal gradient $dT/dz$ of 30 °C/km (Tanaka et al., 2004) and a representative crustal rock density ($\rho$) of 2700 kg/m$^3$. Assumed depth ($z$) conditions were $z$ = 7, 8, 10, 13, 18, 24, and 30 km, with $T = z \cdot dT/dz$ and $P_c = \rho g z$, with $g$ the gravitational acceleration (9.8 m/s$^2$). The name of each experiment and the retrieved sample is denoted as Z#, where # represents the assumed depth (e.g., Z10 refers to an experiment simulating a 10 km depth condition with $T$ = 300 °C, $P_c$ = 265 MPa). We also conducted a hydrostatic experiment at 300 °C and 265 MPa, named "Z10H." All experiments were performed at room-dry conditions. A list of all experiments and conditions employed is given in Table 1.

During an experiment, the axial displacement rate was sequentially stepped between 1 μm/s and 0.1 μm/s to assess the rate dependence of shear stress. Taking a 1 mm thick sample layer, these displacement rates correspond to shear strain rates of $1.4 \times 10^{-3}$ /s and $1.4 \times 10^{-4}$ /s, respectively.

2.2.2. Data correction

In a saw-cut experiment, the load-bearing surface area decreases with increasing axial displacement (Peč, 2014; Peč et al., 2016). Thus, the shear stress was calculated by estimating the area of the piston overlap, similar to Peč (2014). Besides, in calculating the shear strain, we assumed that the sample layer continuously compacted as loading progressed (Peč, 2014).

The initial and final thicknesses of the sample layer are defined as the thickness perpendicular to the pre-cut surface of the shear pistons, and the hydrostatic pressure experiment (Z10H) can estimate the initial thickness. However, a portion of the sample was spilled when the sample was encapsulated in the jacket for experiment Z10H, and the spilled amount was not measured. Thus, the initial thickness was assumed to be 1 mm for all the experiments from the calculation using the density values of quartz and albite in Robie et al. (1967). We verified the initial thickness value by considering the porosity in an area of sample Z10H (~ 30%). A detailed calculation is described in Text S2. Since the amounts of sample and the applied axial displacements are approximately the same in all the deformation experiments, this assumption does not significantly affect strain and strain rate estimates between the experiments. The final thickness was measured from scanning electron micrographs taken from post-mortem sections and resin mounts prepared parallel to the axial loading direction, except for sample Z24, whose final thickness was assumed from the other experiments because the pistons were separated. The final thicknesses of all the experiments are included in Table 1. We plotted the shear stress against shear strain using a smoothing function with a moving average of 30 data points (i.e., 58 seconds).

## 2.3. Microstructural observations and image analyses

The recovered samples were cut into halves along a plane with directions parallel to the loading axis and the shear displacement. We then gave each sample a high-grade polish, as described in Text S3.

Microstructural observations were carried out using a polarizing light microscope (Nikon ECLIPSE C*i* POL at Tohoku University), a Keyence incident light microscope (Keyence VHX-2000 digital microscope at GSJ-Lab of AIST), and scanning electron microscopes (a field-emission scanning electron microscope; FE-SEM (JEOL JSM-7001F at Tohoku University) and a W-tip scanning electron microscope (HITACHI SU3500 at GSJ-Lab of AIST)). The loose fragments were observed with the Keyence microscope and the HITACHI microscope. The FE-SEM has an energy-dispersive spectrometer (EDS), which we used for semi-quantitative element compositional analyses. We also performed Electron backscatter diffraction (EBSD) analyses for samples Z8, Z10, and Z18, although it turned out that a clear Kikuchi band pattern could not be obtained. Therefore, we do not show the results of EBSD here. In fact, similar difficulty in obtaining EBSD data for extremely fine-grained quartz-feldspar aggregates was reported by Peč & Al Nasser (2021). Detailed observational procedures and conditions are explained in Text S3.

Furthermore, we performed transmission electron microscopy (TEM) for samples Z24 and Z30 (Talos F200X at Utrecht University). Using a focused ion beam scanning electron microscope (FIB-SEM, Helios Nanolab G3 at Utrecht University), we cut sections parallel to the half-cut plane so that we could observe the simulated shear zone from directly above. For sample Z24, we also cut another section perpendicular to the half-cut plane, that corresponded to observing shear in the front-back direction.

To visualize the geometry of the phase distribution of quartz and albite, we performed image analyses with the software Fiji (Schindelin et al., 2012). The overview of the analysis procedure is shown in Fig. S2. In this study, we used an image where an element map of Al is overlaid on a BSE image to distinguish mineral phases. As explained in Text S4, we further compared mineral phase distribution maps obtained in the way of this study to the outcomes produced in another way that uses the element map and the BSE image separately (Fig. S3).

Here, we briefly explain the workflow. First, the overlaid image was split into three grayscale images of R, G, and B values. By utilizing the differences in grayscale values between the pixels of the Al plots (mostly albite grains), impurities, surrounding matrix (mostly quartz grains), and cracks (Fig. S2), we obtained binary images showing albite distributions and quartz distributions. We further trained these binary images ("phase plotting maps") using "Trainable Weka Segmentation" (Arganda-Carreras et al., 2017), which is a plugin of Fiji, and obtained probability maps that show mineral-phase distributions. A detailed explanation of the procedure is given in Text S5, and the trained regions are shown in Fig. S4.

Finally, we analyzed the representative phase geometries of different minerals in samples Z18, Z24, and Z30, using the autocorrelation function (ACF). Peč et al. (2016) used ACF to elucidate the shape and connectivity of the slip zone, where grain boundaries were invisible. The situation is similar in this study because grains constituting the sample layer are too small to distinguish individually.

ACF enables us to semi-quantitatively compare the direction and the extent of elongation of each phase at different experimental conditions. The macros for calculating ACF were obtained from https://github.com/kilir/Jazy_macros. The calculation results were displayed as contour maps in the procedure explained in Text S6. The ACF contour maps represent the extent of an overlapping of the analyzed phases when displaced in that direction from the center of each 256-pixel square (i.e., 78 $\mu m^2$). In other words, they indicate how far the phase is distributed in that direction. Thus, the area close to the center shows the local phase shape near the center, while the area far from the center shows the overall phase shape. A detailed explanation of the ACF analysis is provided by Heilbronner (1992). We further set thresholds to the values on the contour maps and compared the shapes of the contour line. We chose the $10^{th}$ bin contour as an inner contour by setting a threshold of 96 to 255. Similarly, we chose the $12^{th}$ bin contour as an outer contour by setting a threshold to the range of 64 to 255. An ellipse was fit to each inner and outer contour line, and the ellipse aspect ratio and orientation of the major axis to the shearing direction were used to describe the shapes of the fit ellipses. All the directions of the major axis are described as a degree measured in the counterclockwise direction from the shear direction.

## 3. Results

### 3.1. Mechanical Data

Fig. 2 shows the stress-strain curves of all experiments. Experiments Z7, Z8, Z10, Z13, and Z18 show strain hardening throughout the experiments, reaching shear strains ($\gamma$) of 3.4-5.0. With increasing simulated depths (i.e., increasing $T$ and $P_c$), the maximum shear stress increased from 710 MPa at $\gamma$ = 4.0 in Z7 to 1300 MPa at $\gamma$ = 3.4 in Z18. Unloading after reaching maximum shear strain went smoothly in most cases, except for experiment Z13, which showed a drop in raw axial stress value by 610 MPa when the temperature was reduced to 50 °C. Experiments Z24 and Z30 show strain weakening beyond $\gamma$ = 2.3 and 0.5, respectively. In experiment Z24, the shear stress ($\tau$) increases to 1280 MPa at $\gamma$ of 2.3, then decreases to 1050 MPa towards the final $\gamma$ of 4.8. In all experiments, $\tau$ increases with increasing strain rate, implying velocity-strengthening behavior. Experiment Z30 underwent much larger strain weakening beyond $\gamma$ = 0.5, from a maximum shear stress of 720 MPa to a final value of 270 MPa at $\gamma$ ~ 5.6. When the $\gamma$ reached ~ 0.4 in experiment Z30, the gear came off the motor shaft, and it stopped rotating for 15-20 minutes, during which time the axial loading stopped. During this period, the temperature showed no specific change except for the variation of the value that is the same extent as observed during the experiment (901 ± 6 °C). Meanwhile, the confining pressure showed a slight increase of ~ 10 MPa, and the axial stress showed a reduction of ~ 200 MPa (Fig. S5a). We removed the data of axial stress and displacement for around this period, as explained in Text S7, to calculate the shear stress $\tau$ and the shear strain $\gamma$ (Fig. S5a).

3.2. Micro- and nano-structures

The stitched SEM images capture crack distributions in samples Z7, Z8, Z10, Z13, and Z18. Using the terminology of Logan et al. (1979), all these samples show cracks in the $R_1$ direction (Fig. 3). The $R_1$ cracks penetrate the entire sample layer and often connect to the cracks overprinting the sample-piston boundary ("*B*" in Fig. 3). Optical microscope analyses of thin sections prepared from samples Z7, Z8, Z10, Z13, and Z18 revealed the presence of relatively fine grains of ~ 10 μm or smaller in size with scattered grains of a tens of micrometers in size (Figs. 4a, 4c, 4e, 5a, and 5c). SEM observations confirmed the presence of submicron-order sized grains (Figs. 4b, 4d, 4f, 5b and 5d). In samples Z7, Z8, and Z10, both quartz and albite grains are mostly comminuted (Fig. 4). Sample Z8 exhibits a coarse quartz grain with cracks extending radially from the center (blue circle in Fig. 4c). In sample Z10, we observed coarse grains (~ 50 μm in size) with pointy-tip shapes (red arrows in Fig. 4e), and elongated fine grains (yellow arrows in Fig. 4e) with aspect ratio ~ 6 and oriented ~ 50° to the shear direction. Samples Z13 and Z18 more frequently exhibit elongated grains (yellow arrows in Figs. 5a and 5c) than samples Z7, Z8, and Z10. Some grains are elongated to an aspect ratio ~ 5 and are oriented 10°-40° to the shear direction (Figs. 5a and 5c). In sample Z13, a large albite grain is elongated at its tips and divided into segments with different extinction angles (green arrow and insert in Fig. 5a). SEM analysis of samples Z8, Z10, Z13, and Z18 show quartz grains with abundant intragranular cracks, which break up original grains into rectangular-shaped fragments 1-3 μm in size (Figs. 4d, 4f, 5b, and 5d). Compared to the quartz grains, albite grains exhibit fewer intragranular cracks, and instead, they show pores of smaller than 1 μm in size (Figs. 4d, 4f, 5b, and 5d). In samples Z13 and Z18 (Figs. 5b and 5d) the pores in albite grains appear to be fewer than in samples Z8 and Z10 (Figs. 4d and 4f). In sample Z18, the boundaries between quartz and albite grains are touching, and even the individual quartz fragments are touching (Fig. 5d).

Furthermore, in sample Z13, highly reflectant surfaces with striations aligned sub-parallel to the shear direction are visible on the surface of the recovered sample (Figs. 5e and 5f). SEM imaging revealed that the striated surfaces are composed of 200-300 nm wide fibers (Fig. 5g).

The sample recovered from experiment Z24 was separated into upper and lower parts (Figs. 3f and 3g). Optical microscope investigation showed zones of localized shear deformation, with porphyroclasts and elongated grains (Fig. 6a). Each porphyroclast has two tails extending from a core of tens of micrometers in size (yellow arrows and lines in Fig. 6b). A large core of quartz is surrounded by rounded grains smaller than 10 μm in size (white arrows in Fig. 6b). Elongated grains are oriented ~ 20° to the shear direction, with an aspect ratio in the range 4-10 (Fig. 6a). One grain is observed which has elongated to ~ 150 μm with an aspect ratio of ~ 10 (Fig. 6c). Near this elongated grain an angular grain of ~ 50 μm is observed, which has a sharp tip facing the bent part of the elongated grain (yellow arrow in Fig. 6c). SEM observations showed that cracks are concentrated in the quartz domains (Fig. 6d). On the other hand, the albite domains are characterized by few pores and an apparent homogeneity (Figs. 6d, 6e, and 6f). They often include elongated structures which are relatively bright when observed using BSE SEM (green arrows in Fig. 6e). One of such bright structures has an aspect ratio of ~ 10 and extends parallel to the shear direction. The quartz domains show fragments having dimensions of a few microns. The outlines of the quartz domains indicate elongation in the direction of 10°-20° with respect to the shear direction (yellow lines in Figs. 6d). The albite domains are also extended in the direction of either parallel to or tilted at a low angle (~ 20°) to the shear direction (Fig. 6d).

TEM observation revealed that both sections of sample Z24 consist mainly of albite domains (Figs. 7a, 7b, and 8a). Both sections clearly exhibit extended structures (Figs. 7b and 8b), which are elongated in the direction of tilted low angle (~ 20°) to the shear direction in the horizontally-cut section (Fig. 7c). These extended structures include nanometer-sized grains as small as 20 nm (Figs. 7d, 8b, and 8c). Although the mineral phase boundary is indistinct in an HAADF (High Angle Annular Dark Field) image (Fig. 7e), EDS maps show a clear boundary between albite and quartz distributions (Figs. 7f-7h). The quartz domains are more fractured than the adjacent albite domain, and even include grains as small as 40 nm in size (Fig. 7e). The diffraction patterns observed in the elongated region (Fig. 8d) exhibit rings with spots, characteristic of a polycrystalline structure (Fig. 8e). In areas with fewer grains within this region (Fig. 8d), the brightness of the diffraction spots diminishes (Fig. 8f).

Optical microscope investigation of sample Z30 revealed a structure dominated by rounded fine grains, some of which are smaller than 10 μm in size, which are surrounded by coarse grains of 50-100 μm in size (Figs. 9a-c). Some grains are elongated to an aspect ratio of 3-15, with their long axes oriented 10°-40° to the shear direction (yellow arrows in Figs. 9a-c). Upon closer examination, boundary shear zones, where grain size is reduced to ~ 5 μm, are observed near the sample-piston boundary (Fig. 9c). Within the boundary shear some grains are elongated with the largest aspect ratio of ~ 15. The grains in the boundary shear zone are aligned in an orientation ~ 10° to the shear direction. SEM imaging confirmed the presence of quartz domains consisting of polygonal grains 200-900 nm in size (Figs. 9d, 9e, and 9f), and albite domains having similar characteristics to those in sample Z24 (Figs. 9d and 9f). The albite domains are elongated in a direction sub-parallel to the shear direction (Fig. 9d). The polygonal grains have a foam texture with 120° dihedral angles at triple junctions (yellow circles in Fig. 9e). Between the polygonal quartz grains, nanometer-sized pores 100-700 nm in size are observed (blue arrows in Fig. 9d). TEM observations show that the whole section consists of nanometer-sized grains (60-210 nm for albite grains, and 230-550 nm for quartz grains). EDS maps show that albite and quartz phases extend parallel with each other (Figs. 10c-10e). Diffraction patterns indicate that both mineral phases are polycrystalline (Figs. 10f and 10g). The bright field TEM image shows crystal defects in quartz grains (red arrows in Fig. 10h), and pores between the grains as indicated by their bright rims (Fig. 10h).

### 3.3. ACF analyses

We summarized the binary images used for the ACF analyses in Fig. 11. Among the samples, sample Z24 exhibits the strongest elongation in both quartz and albite images, and the outlines of the quartz phase in sample Z24 suggest the presence of elongated domains oriented 10°-20° to the shear direction (red arrows in Fig. 11f).

Fig. 12 shows the resulting contour maps. The contour lines are elliptical in the pixel value range of 80-255 (bins 1-11) except for sample Z18 where they are tapered away from the center in both quartz and albite maps. The contour lines of the value range 64-79 (the 12th bin from the highest) are elliptical (Figs. 12c and 12f) in sample Z30, parallelogram-shaped (Figs. 12b and 12e) in sample Z24, and tapered in two directions (Figs. 12a and 12d) in sample Z18. In the maps of sample Z24, the diagonals of the parallelogram-shaped contour lines are directed to ~ 20° and ~ 150° to the shear direction for both mineral phases. In the maps of sample Z18, the taper directions are ~ 20° and ~ 160° in the albite map while ~ 20° and ~ 140° in the quartz map, respectively to the shear direction.

More quantitative comparisons were made using the fit ellipse. Table 2 summarizes the aspect ratio and the major axis direction values of the fit ellipses. The aspect ratio of the ellipse fit to the inner contour is the largest in sample Z24, with 2.4 for the albite contour map and 2.3 for the quartz contour map. The aspect ratios between samples Z18 and Z30 are almost the same, with ~ 1.7 for the albite contour maps and 1.6-1.7 for the quartz contour maps. The major axis directions are nearly the same in the albite contour maps of the three samples (13° in Z18 and Z24, and 10° in Z30), although those in the quartz contour maps differ significantly between the samples (7° in Z18, 13° in Z24, and 4° in Z30).

Meanwhile, the aspect ratio of the ellipse fit to the outer contour is also the largest in sample Z24 for both minerals, with 2.2 in the albite contour map and 2.3 in the quartz contour map. Unlike the inner contours, the aspect ratios in the maps of sample Z30 are slightly larger than those in the maps of sample Z18 for both minerals (i.e., ~ 1.9 in Z30 while ~ 1.6 in Z18). The outer contour in the map of sample Z18 for both minerals tapers in two directions, as mentioned above. The major axis orientation of the fit ellipse measured 172° in the albite contour map and 176° in the quartz contour map, and thus fall between the two taper directions. The major axis orientation of the outer contours in the maps of sample Z24 measured ~ 5°, and the inner contours 13°, hence more sub-parallel to the shear direction than that of the inner contours. For the outer contours in the maps of sample Z30, the major axis directions are almost identical between the inner and the outer contours in the albite contour map (~ 10°). They differ in the quartz contour map (4° for the inner contour and 10° for the outer contour).

## 4. Discussion

The discussion will be divided into two parts, depending on the scale. We will start from the deformation mechanisms at a macroscopic level, based on mechanical data. We show the ACF analysis captured the transition in the dominant deformation mechanism with depth. The other part focuses more on a microscopic level. Possible formation mechanisms of nano-grains and polygonal-shaped grains will be discussed. Finally, we will combine the macroscopic and microscopic levels and envisage a laboratory-obtained strength profile of the continental upper crust.

### 4.1. Macroscopic level shear zones

#### 4.1.1. Depth-dependent strength

To compare mechanical results between the experiments, we calculated the friction coefficient ($\mu$) using values of the shear stress and the confining pressure. The average of the raw data of confining pressure from the entire range of an experiment gave the confining pressure value used in the calculation. Fig. 13 shows values of the shear strain and the obtained friction coefficient smoothed with a moving average of 30 in the same way as the shear stress-shear strain curves (Fig. 2).

In experiments Z7, Z8, Z10, Z13, and Z18, the friction coefficient $\mu$ converges to 0.7-0.8 with increasing $\mu$ (Fig. 13), consistent with Byerlee's friction law (i.e., 0.6-0.85, Byerlee, 1978). By contrast, $\mu$ becomes 0.58 at $\gamma$ = 4.8 in experiment Z24 and 0.23 at $\gamma$ = 5.6 in experiment Z30 (Fig. 13). In experiment Z24, $\mu$ slightly decreases towards the final value after reaching a peak value of 0.63 at $\gamma \sim$ 2.3, while in experiment Z30, a peak value of 0.45 is reached at $\gamma \sim$ 0.5, followed by a decrease to $\mu$ = 0.23 towards the final shear strain of 5.6. The final value of $\mu$ in experiment Z30 is much smaller than the friction coefficient value expected from Byerlee's law. These friction coefficient values suggest a change in deformation mechanisms with increasing temperature and confining pressure in this study. Besides, the shear stress-shear strain curves (Fig. 2) show that experiments Z7 to Z18 (carried out in the temperature and pressure range of respectively 210-540 °C and 185-477 MPa), show strain hardening while experiments Z24 and Z30 show strain weakening (Fig. 2). Moreover, the degree of weakening is greater in experiment Z30 than in experiment Z24. Based on the mechanical results, it is reasonable to suggest that the experimental conditions in this study spanned from brittle to fully plastic transition in quartz-albite aggregates.

#### 4.1.2. Transition in deformation mechanisms highlighted by ACF analysis

Turning now to the deformation mechanisms controlling strain weakening, we focus on a comparison between samples Z18, Z24, and Z30. Samples Z24 and Z30 show strain-weakening behavior, while sample Z18 shows dominant frictional behavior. We chose sample Z18 to represent the samples showing frictional behavior because temperature and confining pressure conditions are the highest among these samples, and hence, the closest to those employed in experiments Z24 and

Z30.

The similarity in the major axis directions of the inner contour in all three albite maps suggests that the albite domains are locally elongated to the maximum stretching direction. Among the samples, sample Z24 shows the largest aspect ratios for both mineral phases and for both inner and outer contours, suggesting that the elongation was the strongest in sample Z24. The major axis directions of sample Z24 suggest that the entire sample layer is stretched and/or aligned in the same direction regardless of the mineral phases.

Meanwhile, samples Z18 and Z30 show almost identical aspect ratios for the inner contour of both mineral phases. As for the outer contour, the aspect ratios of both mineral phases are slightly larger in sample Z30 than in sample Z18. The comparisons between sample Z18 and sample Z30 indicate that the local elongation is to the same extent between the two samples while sample Z30 shows a stronger alignment than sample Z18. As for sample Z30, the major axis directions represent that, for the albite phase, both individual domains and the entire layers are elongated in the maximum stretching direction. On the other hand, for the quartz phase, individual domains are not elongated to the maximum stretching direction, but their alignments are still oriented to the maximum stretching direction.

Furthermore, the shapes of the outer contours suggest a transition in deformation mechanisms with simulated depth. The outer contours of samples Z18 and Z24 have corners (Fig. 12) in directions similar to that of an $R_1$ shear (i.e., ~ 165°; Logan et al., 1979). Fig. 3 shows that sample Z18 exhibits cracks that are oriented to 140° ~ 150° and penetrate through the sample layer (Fig. 3e), while sample

Z24 was separated into upper and lower parts by a surface oriented to ~ 170° (Figs. 3f and 3g), both with respect to the shear direction. These results suggest an operation of frictional slip along the $R_1$-shear direction. Since the other bulges of the outer contour of samples Z18 and Z24 are oriented to the maximum stretching direction (Figs. 12a, 12b, 12d, and 12e), the shapes of the outer contours represent that both the frictional slip along the $R_1$-shear direction and the grain elongation in the maximum stretching direction operated in samples Z18 and Z24. On the other hand, the outer contours of sample Z30 do not have such bulges (i.e., inner and outer contours match; Figs. 12c and 12f), indicating that frictional slip was negligible in sample Z30.

## 4.2 Microscopic level shear zones

### 4.2.1. Outline of a microstructural change

In samples Z7, Z8, Z10, Z13, and Z18, both quartz and albite grains are comminuted (Figs. 4 and 5), suggesting that the imposed shear strain was mainly accommodated by brittle grain failure (fracture) and grain translation. Meanwhile, samples Z24 and Z30 show nano-grains, as well as quartz porphyroclasts and polygonal-shaped grains (Figs. 6 and 9). Despite such a distinctive transition from fracture to flow with depth, optical and electron microscope observations revealed that plastic deformation processes become more prominent, even when mechanical behavior is still consistent with Byerlee’s law. Specifically, in samples Z7, Z8, and Z10, most grains are comminuted, while samples Z13 and Z18 showed more widespread elongated grains suggestive of plastic flow. Moreover, intragranular pores in albite become less abundant in samples Z13 and Z18 compared to samples Z8 and Z10. Elongated grains are oriented 10°-40° with respect to the shear direction, which is sub-parallel to the maximum stretching direction (cf. an S shear plane, see Figs. 5a and 5c and Berthé et al., 1979).

Interestingly, sample Z13 exhibits striations on the surface of the sample-piston boundary (Figs. 5e-g). Similar striations were found on the sheared antigorite samples retrieved from shear deformation experiments by Brantut et al. (2016) carried out at room temperature, confining pressures of 30 and 95 MPa, while employing a net slip rate of 8 μm/s. Using SEM observations, the authors showed that the striations comprise highly porous zones, which formed due to rapid slip events upon unstable sliding observed in the experiments (“stick-slip” behavior). The rupture speeds in their experiments were estimated 3.5-6.5 km/s. Meanwhile, experiment Z13 in this study showed an abrupt drop of the raw axial stress value by 610 MPa during the quenching stage. The apparatus stiffness during this period was 4 GPa/mm, that is larger than the Griggs apparatus stiffness values summarized in Burdette & Hirth (2020). Considering the apparatus as a spring, the relatively large stiffness represents an instantaneous high-speed sliding. Following Brantut et al. (2016), we interpret the striations in sample Z13 to have formed in this unstable slip event.

4.2.2. Nano-grains

Samples Z24 and Z30 exhibit areas with few pores and apparent homogeneity (Figs. 6 and 9). They resemble the areas of the localized strain, or the slip zones, formed in the quartz-feldspar aggregates in Peč & Al Nasser (2021). They inferred them to be a similar material as nanocrystalline to partially amorphous material, based on SEM and TEM observations in previous studies (Peč et al., 2012, 2016; Marti et al., 2017, 2020). On the other hand, TEM observations in this study revealed that these areas contain almost no amorphous materials, and instead are fully nanocrystalline (Figs. 7, 8, and 10). Below, we discuss the roles of nano-grains in controlling the bulk mechanical behavior.

Nanocrystalline materials are reported in both natural and experimentally simulated slip zones (see Verberne et al., 2019 for a review). Green et al. (2015) concluded that nanograins in fault zones cause extreme weakening in high-speed frictional sliding. Sun & Peč (2021) pointed out that nanocrystalline fault rocks have intrinsically low viscosity, potentially leading to a seismic unstable slip by forming a kinematically favorable failure plane. The nanogranular areas in this study extend either parallel or at a low angle to the shear direction, and almost in the direction that is favorable for efficiently accumulating strain (Peč et al., 2016; Peč & Al Nasser, 2021). Considering the strain-weakening behavior of experiments Z24 and Z30, the interconnected nanogranular domains are expected to have controlled the bulk mechanical behaviors in samples Z24 and Z30.

The observed nano-grains are tens of nanometers in size at the smallest for both minerals in sample Z24 (quartz: 40 nm and albite: 20 nm). Similarly, sample Z30 exhibits an albite grain of 60 nm in size, although the smallest quartz grain is relatively larger with a grain size of ~ 200 nm. Sammis & Ben-Zion (2008) showed that a critical particle size $d_{crit}$ below which fracture becomes impossible is given by

$$d_{crit} = \frac{32}{3}\left(\frac{K_{Ic}}{Y}\right)^2, \quad (1)$$

where $Y$ is the compressive yield stress and $K_{Ic}$ is the critical stress intensity factor for mode I rupture, which was defined as

$$K_{Ic}^2 = EG_c, \quad (2)$$

using Young's modulus $E$ and the fracture energy $G_c$. Using these relationships, we calculated a critical particle size which can be produced by fracture. The value of Young's modulus varies at different temperatures, as 104.7 GPa at 720 °C and 107.9 GPa at 900 °C for quartz (Pabst & Gregorová, 2013; calculated from measured values at room pressure in Lakshtanov et al., 2007). For albite, we applied $E$ of 86.5 GPa for both temperatures, which was calculated by taking an average of the simulation- and experimentally derived values at room temperature reported in Pabst et al. (2015). The values of $G_c$ at room temperature were measured to be 0.410-1.030 $Jm^{-2}$ for quartz (Brace & Walsh, 1962), and 7.770 $Jm^{-2}$ for orthoclase (Brace & Walsh, 1962), which we assumed to be similar to that of albite. Considering the maximum differential stress during experiments, $d_{crit}$ of experiment Z24 becomes 70-176 nm for quartz and 1.1 μm for albite, while those of experiment Z30 are 230-570 nm for quartz and 3.4 μm for albite. Thus, the observed minimum grain sizes are smaller than the calculated $d_{crit}$.

Previous studies suggested that nano-grains could be formed by several processes including comminution, chemical-reaction, and plastic deformation. Peč et al. (2016) inferred that the nanocrystalline to amorphous materials in experimentally simulated granitoid fault rocks were formed through an amorphization process of comminution as proposed by Yund et al. (1990). However, the above calculation showed that the observed nano-grains in this study are smaller than the critical particle size produced by fracture, implying that processes other than comminution took place to produce the nano-grains in samples Z24 and Z30. Meanwhile, chemical reaction is pointed out by Ohl et al. (2020), where they proposed that grains smaller than 50 nm in size in natural carbonate rocks were produced by pseudomorphic replacement of calcite. Nevertheless, chemical reactions between quartz and feldspar could be less possible, due to the robust crystal lattice structures of quartz that is composed of covalent bonds. The unlike possibility of the chemical alteration is supported by our EDS analyses. Fig. S6 shows that the ratios of Na: Al: Si does not change significantly between the SEM analyses and TEM analyses. Considering that the SEM analyses represent the chemical composition of a relatively large area, while the TEM analyses show the chemical composition of a relatively small area, this result suggests that the local variation in chemical composition is unlikely in samples Z24 and Z30. From the above discussion, we infer that plastic deformation was involved in forming nano-grains in this study.

Grain boundary sliding has been pointed out to play an important role in deformation of nanocrystalline metals (Lu et al., 2000; Kumar et al., 2003). It has also been regarded as a key deformation mechanism in geological context. Green et al. (2015) suggested that nanocrystalline fault materials flow by grain boundary sliding. Ohl et al. (2021) inferred that dislocation activity accompanied by grain boundary sliding may have formed the nanostructures in natural carbonate fault rocks. As for quartz aggregates, Fukuda et al. (2018) proposed a flow law considering the grain boundary sliding. Furthermore, Marti et al. (2018) suggested that a crystallographic preferred orientation (CPO) observed in extremely fine-grained plagioclase-pyroxene mixtures was a resultant of simultaneously operated dissolution-precipitation creep and grain boundary sliding. Because samples Z24 and Z30 in this study were easily damaged by the TEM electron beam, we could not obtain information of CPOs. Nevertheless, the existence of nano-grains in these samples implies a potential that grain boundary sliding was the controlling deformation mechanism. Besides, experiments Z24 and Z30 in this study showed strain weakening (Fig. 2). Précigout et al. (2007) proposed that grain boundary sliding plays a significant role in the weakening of the continental lithosphere. Similarly, Warren & Hirth (2006) proposed that grain boundary sliding, coupled with grain size reduction and second phase pinning, results in weakening and strain localization in the oceanic lithosphere. Therefore, mechanical results also support that grain boundary sliding occurred in sample Z24 and Z30. Although strain weakening could be caused by other processes including dynamic recrystallization (Cross & Skemer, 2019), CPO development (Fan et al., 2021), and phase mixing (Cross & Skemer, 2017; Wiesman et al., 2018), the existence of nano-grains raises the possibility of grain boundary sliding, considering the frequent involvement of grain boundary sliding with nanogranular materials in the previous studies (Lu et al., 2000; Kumar et al., 2003; Green et al., 2015; Demurtas et al., 2019; Pozzi et al., 2019; Ohl et al., 2021).

In addition, triple junctions have been considered to be important in controlling properties of nanocrystalline materials (e.g., Palumbo et al., 1990; Ryou et al., 2018). Recently, Yokoyama &

Nagahama (2025) showed a relationship between the volume fractions of triple junctions and rank-1 connection, which was defined as the conditions where geometric compatibility is satisfied across an interface, that is, the difference in the deformation gradients between two adjacent areas is described as a rank-1 matrix (Ball & James, 1987). It was suggested that a higher volume fraction of triple junctions indicates that the strength of the interface is no longer supported by the rank-1 connection (Yokoyama & Nagahama, 2025). In this study, triple junctions are observed in the quartz polygonal grain domains in sample Z30 (Fig. 9e), and pores exist in their neighbors (Fig. 9d). These observations suggest that contacts between grain surfaces were inhibited and hence the rank-1 connection was not satisfied. Such a collapse of the rank-1 connection at the nanoscale level potentially explains the strain-weakening behavior of experiment Z30. Although the sample fragility made it hard to visualize grain boundaries within the nano-grain domains, future analyses focusing on nano-grain boundaries will provide experimental insights into the role of triple junctions on the bulk mechanical behaviors.

#### 4.2.3. Formation of polygonal grains

Experiment Z30 shows a more significant weakening than experiment Z24 (by ~ 450 MPa in experiment Z30 while by ~ 240 MPa in experiment Z24), indicating that a weakening process unique to sample Z30 took place. Such an assumption that a unique weakening process took place in experiment Z30 is also supported by microstructural analysis, as sample Z30 uniquely shows quartz polygonal grains (Figs. 9d, 9e, and 9f) while sample Z24 exhibits aggregates of angular quartz fragments (Figs. 6d and 6e) resembling the lens-shaped fractured aggregates in Peč et al. (2016). Given such differences in microstructure, the process of forming the quartz polygonal grains must be responsible for the significant weakening of experiment Z30.

Possible formation processes of the quartz polygonal grains in sample Z30 are dynamic recrystallization during the experiment and/or static grain growth of nanograins during the quench just after the experiment. In fact, Verberne et al. (2017) suggested that post-shear static grain growth may have played a role in their shear deformation experiments on simulated calcite fault rock. To estimate the potential grain growth in our experiments, we used a grain growth law for quartz (Karato, 2008, p.237 & p.241; Fukuda et al., 2018), which was based on experiments on wet, fine-grained quartz aggregates by Tullis & Yund (1982). Hirth et al. (2001) pointed out that at room-dry, or "as-received" conditions, experimental observations suggest that water is sufficiently included to enhance the creep rates in the flow of quartzite compared to dry conditions, and additional water causes further weakening. Since our experiments were performed at room-dry conditions, using the grain growth law calibrated to experiments performed under wet conditions would yield a maximum estimate of growth of grains in sample Z30. Following Fukuda et al. (2018), we employ values of a constant factor ($k_0$) of 100 $\mu m^2/s$ and an activation energy ($Q_g$) of 80 kJ/mol.

The observed grain sizes in sample Z30 are in the range of 200-900 nm (Figs. 9d, 9e, and 9f). During the quenching of experiment Z30, the temperature decreased from 900 °C to 300 °C in ~ 20 seconds. Therefore, the upper bound of the amount of growth is obtained by assuming that the temperature was constantly 900 °C, while the lower bound of that is obtained by assuming that the temperature was constantly 300 °C. In the former assumption, the observed grain in a size of 900 nm is calculated to be formed from a grain of ~ 500 nm, and a grain smaller than ~ 700-800 nm cannot exist. Meanwhile, in the latter assumption, the grains rarely grow within a time range of 20 seconds. A more realistic assumption can be made by taking an average of the two assumptions, that is, that the temperature was constantly 600 °C during the 20 seconds of quenching. Under 600 °C, the observed grain sizes are calculated to be grown from grains initially 90-900 nm in size (Fig. S7).

Using a grain size piezometer, we now calculate the possible grain sizes that can be formed from dynamic recrystallization in experiment Z30. The mechanisms of dynamic recrystallization in experimentally deformed quartz aggregates are divided into three dislocation creep regimes (Hirth & Tullis, 1992), depending on stress (Stipp & Tullis, 2003). As the final differential stress of experiment Z30 (540 ± 30 MPa) is consistent with those of quartzite and novaculite deformed in regime 1 in Hirth & Tullis (1992), it is likely that the dislocation creep regime 1, that is bulging recrystallization (Stipp et al., 2002), took place. The piezometer for dislocation creep regime 1 of quartz is given by (Post & Tullis, 1999; Stipp & Tullis, 2003)

$$d = \varphi \times \sigma^{-0.61 \pm 0.04}, \quad (3)$$

where $\varphi$ is $10^{1.89 \pm 0.11}$ ($\mu\mathrm{m}/[\mathrm{MPa}]^{-0.61 \pm 0.04}$), $d$ is recrystallized grain size (μm) and $\sigma$ is differential stress (MPa). The peak and final differential stress $\sigma$ of experiment Z30 is 1450 ± 30 MPa and 540 ± 30 MPa, respectively, as each corresponding shear stress $\tau$ is 720 ± 30 MPa and 270 ± 30 MPa, and the shear pistons were cut at 45° to the loading axis. Thus, by applying the Reuss bound, which assumes that the stress acting on the quartz phase is the same as the bulk stress, the recrystallized grain sizes are estimated to be 520-1600 nm at the peak stress and 970-2900 nm at the final stress. The grain sizes just before the quench (90-900 nm) are consistent with the smaller range of the estimation for the peak stress. Besides, even if the effect of the static grain growth is much smaller than the extent estimated above and is negligible, the observed grain sizes (200-900 nm) is consistent with the estimated grain sizes at the peak stress. Therefore, dynamic recrystallization more likely took place at the peak stress and the recrystallized grains were preserved until the end of the experiment, rather than it proceeded towards the end of the experiment with an equilibrium. Nonequilibrium recrystallization grain size with evolving stress was shown for cooling-ramp experiments (Soleymani et al., 2020).

Despite the above estimations, it should be noted that there remains a possibility that local stress buildup produced the polygonal grains even at the end of the experiment. The grooves made on the pre-cut piston surfaces likely caused this stress concentration. Moreover, assuming that Equation 3 is valid for nanometer scale, differential stresses of tens of GPa are required to generate the nano-grains smaller than 100 nm. Although it remains unclear whether the stress could build up to this extent, the piston grooves may have locally enabled the production of nano-grains through dynamic recrystallization. Meanwhile, local stress buildup also results in a smaller critical particle size $d_{crit}$, suggesting that grain size could have been reduced to the nanometer scale of the observed grains by comminution. Nevertheless, at least the quartz polygonal grains were likely formed by recrystallization, since comminution should have produced angular-shaped grains. Future analyses using high-angular-resolution electron backscatter diffraction (HR-EBSD) will be beneficial for highlighting the extent of stress heterogeneity (Wallis et al., 2020; Wiesman et al., 2024).

Furthermore, the extreme grain size reduction to the nanoscale was potentially accomplished by the operation of a Frank-Read source. Once a pair of pinning points in the polycrystalline sample began operating as a Frank-Read source, it would have multiplied dislocations and may have produced nano-grains. Aside from the dynamic recrystallization involving a Frank-Read source, subcritical fragmentation may explain the generation of the nano-grains (Sammis & Ben-Zion, 2008). Nano-grains could locally have grown into polygonal grains, aided by local frictional heat, which was estimated in laboratory experiments (Ortega-Arroyo et al., 2025), as well as through dynamic grain growth (Rofman et al., 2009). Local melting might also be possible considering that the as-is samples likely absorbed water. However, we expect melting to be less effective in the bulk mechanical behavior based on the TEM observations exhibiting the polycrystalline nature. Future experiments with various strains will shed new light how nano-grains are formed and coexist with submicron-sized grains in shear zones.

### 4.3. Implications for crustal strength and future studies

Summarizing the results offers a picture of a depth profile of the upper crust, spanning the depth region where the dominant deformation mechanism changes from fully brittle deformation to fully plastic deformation (Fig. 14). At the shallower depth part, shear stress increases linearly with depth, following Byerlee's friction law. In this region, stress reaches a peak at the final strain, indicating that strain weakening does not occur. The microstructures of the samples deformed under the conditions in this region mostly consist of comminuted grains and suggests a predominance of brittle deformation, although grains are more frequently elongated at greater depths in this region. On the other hand, at the greater depth part of the upper crust, significant strain weakening takes place. Both albite and quartz particles become nanocrystalline, although quartz particles often remain as coarser fragments at a shallower depth. These results suggest that grain boundary sliding at the nanocrystalline areas causes the strain weakening, and hence determines the crustal strength of the brittle-plastic transition zone.

The final shear stress in experiment Z30 is smaller than the Goetze's criterion value (red line in Fig. 14), suggesting that the condition of experiment Z30 is within the plastic region. The curves in Fig. 14 indicate flow stresses of minerals at the faster strain rate ($1.4 \times 10^{-3}$ /s). The flow law parameters are derived from Offerhaus et al. (2001) in Rybacki & Dresen (2004) for albite (purple curve in Fig. 14), and revised Luan & Paterson (1992) in Fukuda & Shimizu (2017) for quartz (blue curve in Fig. 14). For the 50-50 albite-quartz mixture, we used the equations for the two end members, which are the uniform strain rate bound (uniform strain rate curve in Fig. 14) and the uniform stress bound (uniform stress curve in Fig. 14), in Tullis et al. (1991). We substituted $\sigma = 2\tau$ and $\dot{\varepsilon} = \dot{\gamma}/\sqrt{2}$ to the flow laws. The peak shear stress of experiment Z30 is almost equal to the quartz flow stress, implying that plastic flow of quartz controlled the peak stress of the aggregates. This is consistent with the discussion in section 4.2.3 where we suggested the possibility that dynamic recrystallization of quartz took place at the peak stress. Interestingly, the final shear stress of experiment Z30 is consistent with the albite flow stress, and the shear stresses during the strain weakening are the same extent as the flow stresses of the mixture, which are estimated as the value between the two bounds (Tullis et al., 1991). This result may imply that in experiment Z30, the bulk mechanical behavior was controlled by quartz flow at the peak stress, but gradually albite flow became dominant towards the end of the experiment. Similar strength evolution from near uniform strain rate bound to near uniform stress bound with increasing strain was also reported by Cross et al. (2020). Nevertheless, the existence of the nano-grains itself results in reducing the stress as discussed above, and it is unclear to what extent it could be applicable to interpret the relationship between the shear stress and the flow laws as representing the change in the dominant mineral phase with strain. What we would like to emphasize here is that plastic deformation was suggested to be the dominant deformation mechanism under the condition of experiment Z30, from the perspective of the stresses estimated by the flow laws. The coexistence of nano-grains and polygonal grains at the greatest depth part of the brittle plastic transition zone potentially suggests that nano-grains grow into polygonal grains by dislocation activities in the process of dynamic recrystallization. Although nano-grain formation process remains unclear, the coexistence of nano-grains and polygonal grains suggests that grain boundary sliding and dynamic recrystallization concurrently weaken the crustal strength at the greatest depth part, resulting in a significant reduction of the crustal strength there.

Furthermore, the peak shear stresses at the simulated depths of 18 km and 24 km are almost identical and larger than those in the shallower or greater depth regions. This result suggests that the depth where the stress becomes the largest in the upper crust exists in width as shown in Shimamoto & Noda (2014). Here we provide microphysical evidence for the smooth brittle-plastic transition by pointing out the concurrent occurrence of several weakening processes. In this way, we can conclude that this study revealed deformation mechanisms that control the crustal strength at the brittle-plastic transition zone as Fig. 14 clearly shows. Such an experimentally-based series of deformation mechanisms spanning the entire brittle-plastic transition zone is unprecedented for quartz-feldspar aggregates, as represented by the temperature and confining pressure of experiment Z30 (900 °C and 870 MPa) that are higher than those in Peč & Al Nasser (2021) (750 °C and 800 MPa).

It is also supported by several studies on pores that the deformation mechanism at the highest temperature and confining pressure in this study (i.e., experiment Z30) simulates that at the greatest depth part of the brittle-plastic transition zone. Cavitations in experimentally simulated gouges (Rybacki et al., 2008; Verberne et al., 2017) and in natural mylonites (Shigematsu et al., 2004; Yeo et al., 2025) have been considered to trigger catastrophic rupture at the brittle-plastic transition zone (Shigematsu et al., 2004; Verberne et al., 2017; Yeo et al., 2025) to possibly the lower crust (Rybacki et al., 2008). Cavitations are also considered significant for fluid flows in the middle to lower crust (Fusseis et al., 2009; Menegon et al., 2015). In this study, pores are observed between the polygonal grains of sample Z30 (Fig. 9d). This observation suggests the potential for a catastrophic rupture followed by runaway fault slip under further strain, at the depth condition simulated in experiment Z30.

**5. Conclusions**

We performed shear deformation experiments on quartz-albite aggregates with a Griggs-type solid medium apparatus under various temperatures and confining pressures simulating the realistic crustal depth conditions. Mechanical data at the shallower depth conditions (temperatures and confining pressures of 540 °C and 477 MPa or lower) show strain hardening and the peak shear stress increases as the simulated depth becomes greater. Both albite and quartz grains are fractured at these conditions, although a subtle difference between these conditions exists that elongated grains are more often observed at greater depths. These results suggest that brittle deformation dominates at these conditions, which is supported by a convergence of the friction coefficient to 0.7-0.8, following Byerlee's friction coefficient. On the other hand, at a temperature of 720 °C and a confining pressure of 750 MPa, strain weakening behavior is observed, although the peak shear stress remains almost identical to that at 540 °C and 477 MPa. The sample exhibits nanometer-sized grains of both albite and quartz, embedded in the matrix composed of coarser quartz fragments. Furthermore, at the greatest depth condition (a temperature of 900 °C and a confining pressure of 870 MPa), the final shear stress is much lower than all the other conditions. The sample exhibits nanometer-sized albite grains and polygonal-shaped quartz grains. These findings suggest that the extreme grain size reduction to nanometers, as well as dynamic recrystallization at the greatest depth part, causes the strain weakening behavior.

We conclude that grain boundary sliding of nanocrystalline materials controls the peak crustal strength in the brittle-plastic transition zone of the upper crust. At the greatest depth in the brittle-plastic transition zone, nanogranular slip and dynamic recrystallization concurrently reduce fault strength and provide a potential for abrupt failure.

**Acknowledgements**

The authors are thankful for Caleb Holyoke, Andrew Cross, and Jacques Précigout for their thoughtful reviews that improved the manuscript. We thank Maartje Hamers for preparing the FIB-sections for TEM observations. We also acknowledge Masanori Kido for offering the backscattered electron images of samples Z7 and Z18, Sambuddha Dhar for offering the script for drawing the strength profile, and Shun Arai for his help with the experiments. We further acknowledge the Electron Microscopy Centre at Utrecht University for providing the FIB-SEM and the TEM as well as the EPOS-NL MINT facility. This work was supported by The International Joint Graduate Program in Earth and Environmental Sciences, Tohoku University (GP-EES), and JSPS KAKENHI Grant Number JP23KJ0205. O. Plümper was supported by an ERC Starting Grant (nanoEARTH, #852069).

**Data Availability Statement**

Data from the deformation experiments are available in the Mendeley Data repository at https://data.mendeley.com/preview/9s84ny8fmd?a=a442a935-bf3d-4b85-a955-f77700b00012. This is a private peer review link. You can access the data after you log-in.

**Conflict of Interest Statement**

The authors have no conflicts of interest to disclose.

**References From the Supporting Information**

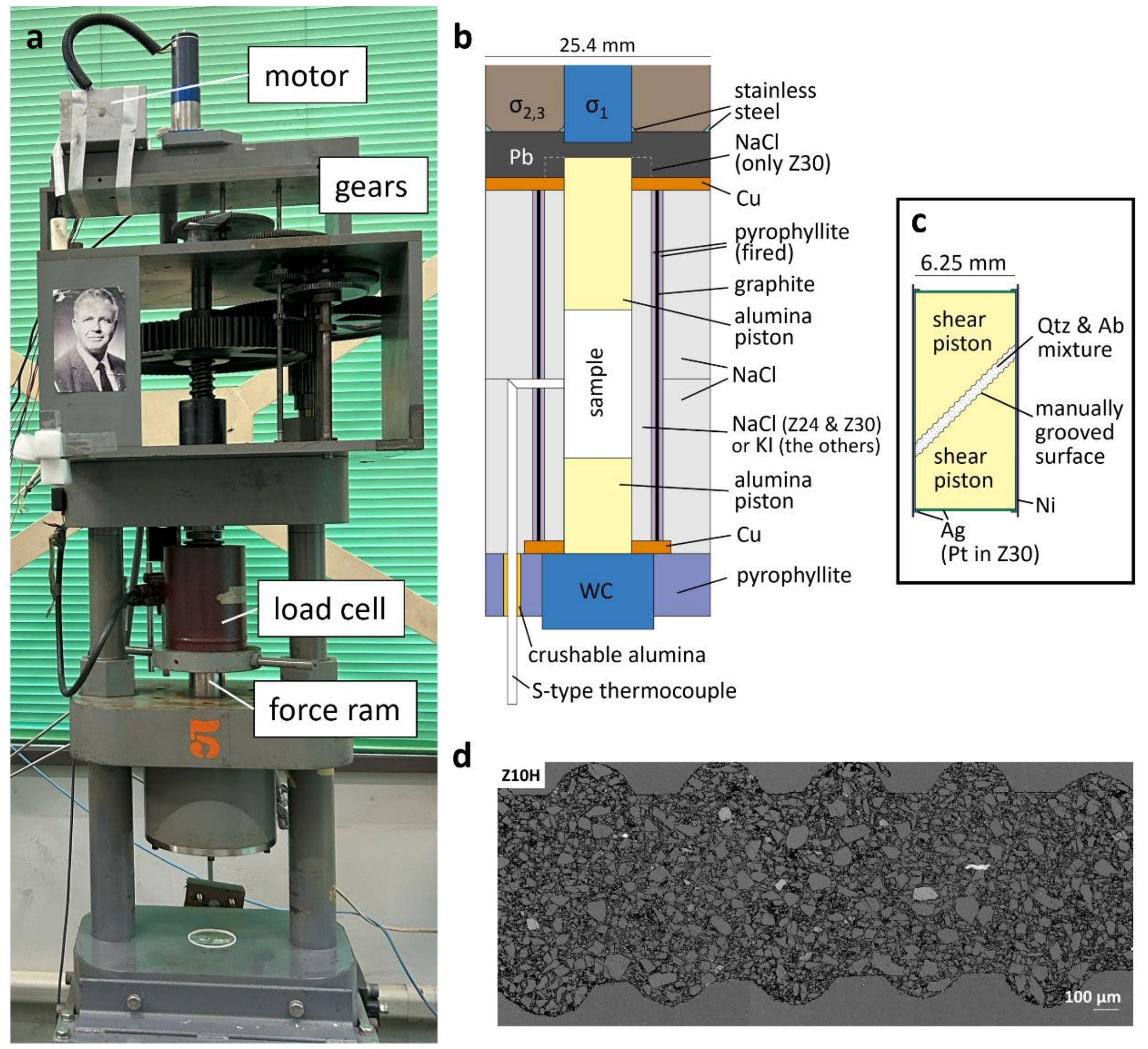

Figure 1. Summarized images of the experimental setup. (a) Photograph of the Griggs-type deformation rig installed at Tohoku University. (b) Schematic cross-section of the sample assembly (modified after Kido et al., 2016). (c) Schematic cross-section of the sample and pistons encapsulated in the jackets. (d) Stitched BSE image showing an area in the hydrostatic experiment (experiment Z10H).

Table 1. Summary of experimental conditions and mechanical data of this study

| Sample name | Temperature $T$ (°C) | Confining pressure $P_c$ (MPa) | Simulated depth $z$ (km) | Final thickness (mm) | Peak shear stress (MPa) | Final shear stress (MPa) | Shear strain at peak shear stress | Final shear strain |
|---|---|---|---|---|---|---|---|---|
| Z7 | 210 | 185 | 7 | 0.763 | 709 | 709 | 4.0 | 4.0 |
| Z8 | 240 | 212 | 8 | 0.740 | 840 | 840 | 4.1 | 4.1 |
| Z10 | 300 | 265 | 10 | 0.760 | 880 | 880 | 4.2 | 4.2 |
| Z13 | 390 | 344 | 13 | 0.609 | 1116 | 1116 | 5.0 | 5.0 |
| Z18 | 540 | 477 | 18 | 0.802 | 1303 | 1303 | 3.4 | 3.4 |
| Z24 | 720 | 750 | 24 | 0.800* | 1279 | 1053 | 2.3 | 4.8 |
| Z30 | 900 | 870 | 30 | 0.560 | 724 | 271 | 0.5 | 5.6 |
| Z10H | 300 | 265 | 10 | 0.713 | — | — | — | — |

*Assumption (not measured).

Axial displacement rates were repeatedly changed in all experiments between 1 μm/s and 0.1 μm/s corresponding to shear strain rates of $1.4 \times 10^{-3}$ /s and $1.4 \times 10^{-4}$ /s, respectively. The symbols "—" denote that no values exist since experiment Z10H was performed under a hydrostatic condition.

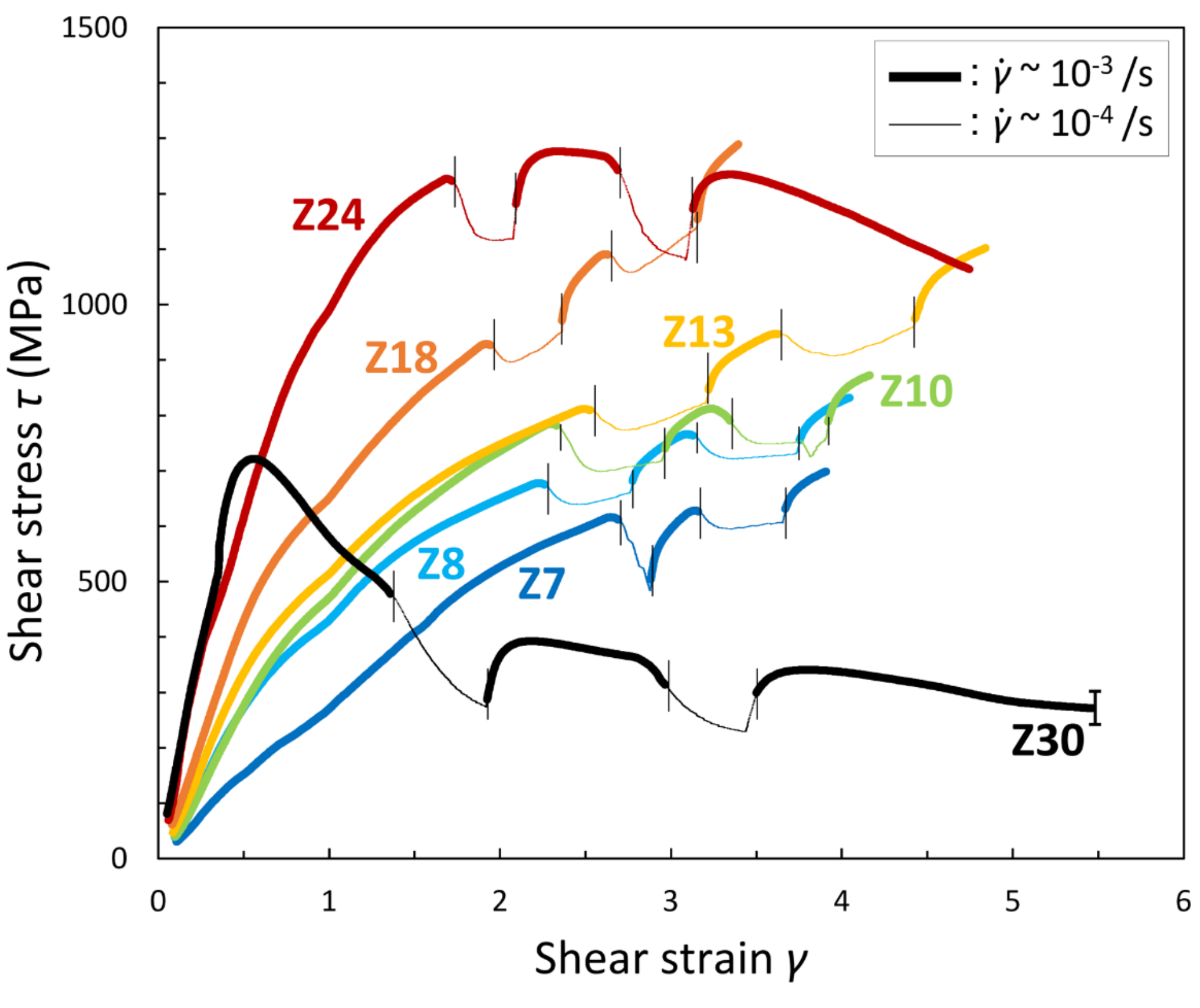

Figure 2. Relationships between shear stress $\tau$ and shear strain $\gamma$ during the experiments. Bold lines: shear strain rate of ~ $10^{-3}$ /s, thin lines: shear strain rate of ~ $10^{-4}$ /s. The lines in blue: Z7, light blue: Z8, light green: Z10, yellow: Z13, orange: Z18, red: Z24, and black: Z30. Moving averages of raw data of $\tau$ and $\gamma$ were calculated with an interval of 30 and plotted. The vertical lines indicate where the strain rate was switched. Since we calculated the moving average, 30 data between each step become an average of the values originated from both strain rate steps. For those data, we regarded that the resulting average values belong to the strain rate step where the greater number of original values came from (i.e., for example, if the average was calculated from 10 data of the faster strain rate step

and 20 data of the slower strain step, the resulting average value was regarded belonging to the slower strain rate step. The middle data, which was generated from 15 data from both strain rate steps, were removed.). Note that the graph of Z30 is discontinuous around the $\gamma \sim 0.4$ because of the accidental cessation of the axial loading. The error bar is shown at the final point of the result of experiment Z30 (± 30 MPa).

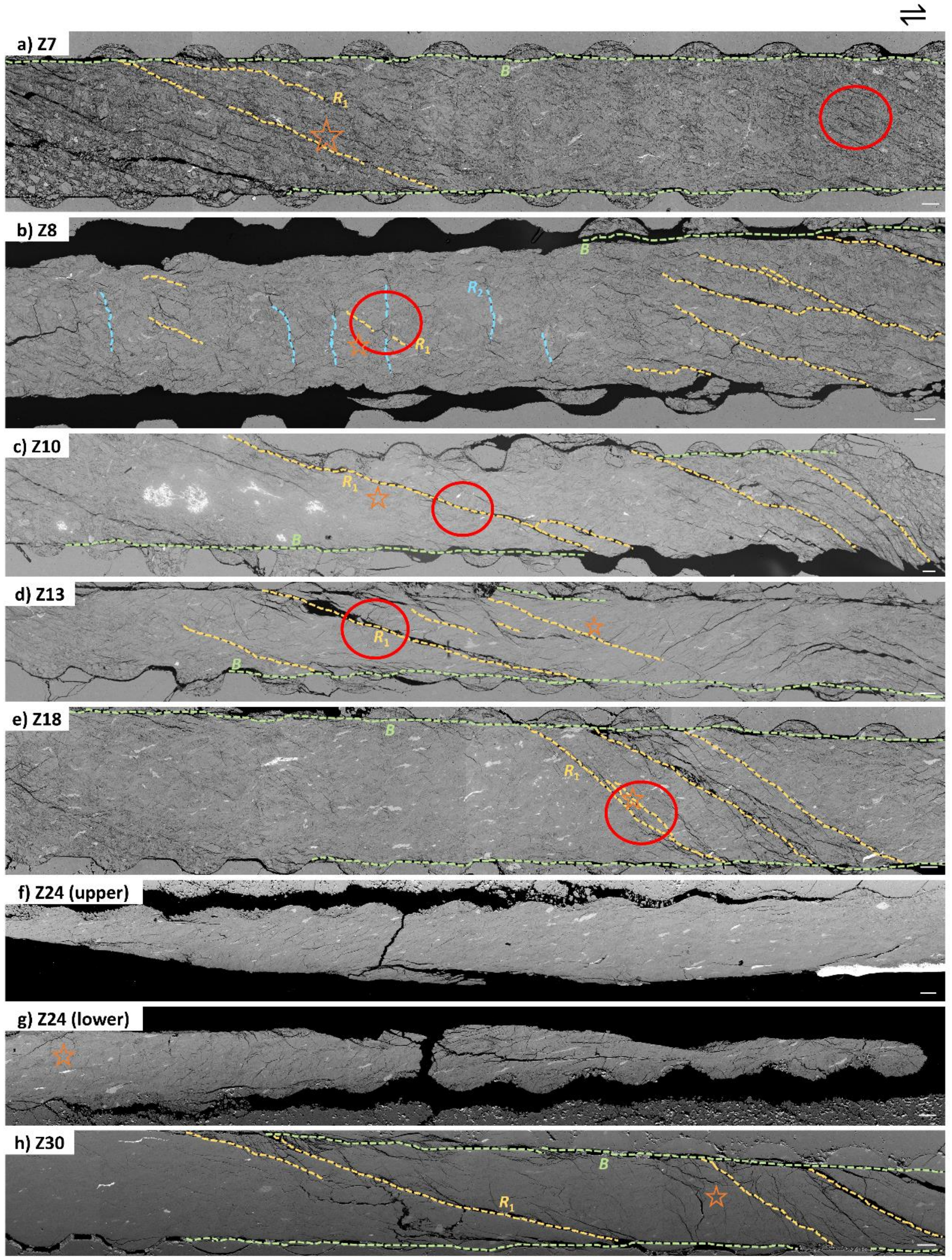

Fig. 3 Furukawa et al. (2025)

Figure 3. Stitched BSE images near the center of sample Z7 (210 °C & 185 MPa; Figure 3a), Z8 (240 °C & 212 MPa; Figure 3b), Z10 (300 °C & 265 MPa; Figure 3c), Z13 (390 °C & 344 MPa; Figure 3d), Z18 (540 °C & 477 MPa; Figure 3e), Z24 (the upper half) (720 °C & 750 MPa; Figure 3f), Z24 (the lower half) (720 °C & 750 MPa; Figure 3g), and Z30 (900 °C & 870 MPa; Figure 3h). All samples from deformation experiments have $R_1$ cracks (yellow dotted lines) as well as the cracks developing at the sample-piston boundary in the direction parallel to the shear direction ($B$; green dotted lines) except for Z24. In addition, Z8 has $R_2$ cracks (blue dotted lines in Figure 3b). Scale bars are 100 μm. All images are the dextral sense of shear. Red circles in Figures 3a to 3e denote the approximate locations where the images of the polarizing light microscope in Figures 4 and 5 were taken. Stars in Figures 3a, 3b, 3c, 3d, 3e, 3g, and 3h denote the approximate locations of SEM images shown in Figures 4, 5, 6, and 9 were taken.

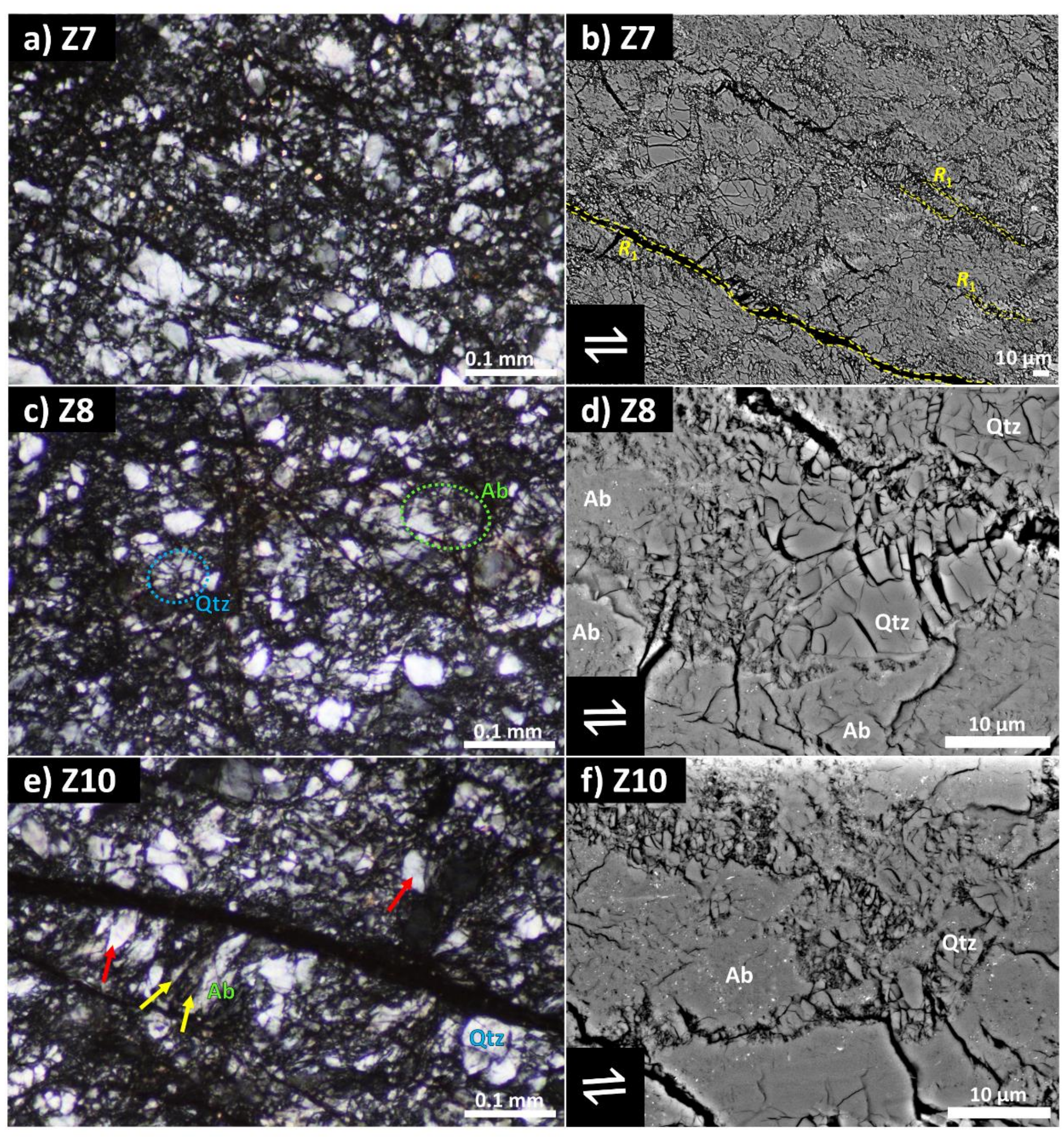

Fig. 4 Furukawa et al. (2025)

Figure 4. Photomicrographs of samples Z7 (210 °C & 185 MPa; a and b), Z8 (240 °C & 212 MPa; c and d), and Z10 (300 °C & 265 MPa; e and f). Figures 4a, 4c, and 4e were taken under crossed polars, while Figures 4b, 4d, and 4f were taken with SEM. (a) Coarse grains are embedded in the matrix composed of finely comminuted grains. (b) Extremely fine grains especially near the $R_1$- cracks (yellow lines). (c) Coarse quartz grain (blue circle) and albite grain (green circle) show differences in the crack geometry. (d) Extremely fine grains between the coarse quartz and albite grains. The coarse quartz grains are separated into rectangular-shaped fragments. (e) Some fine grains are elongated (yellow arrows). Red arrows denote coarse grains with pointy-tip shapes. (f) Extremely fine grains between the coarse quartz and albite grains. The coarse quartz grains consist of rectangular-shaped fragments. The shear sense in Figures 4a, 4c, and 4e are approximately dextral, and those in Figures 4b, 4d, and 4f are shown with white arrows. Qtz: quartz, Ab: albite.

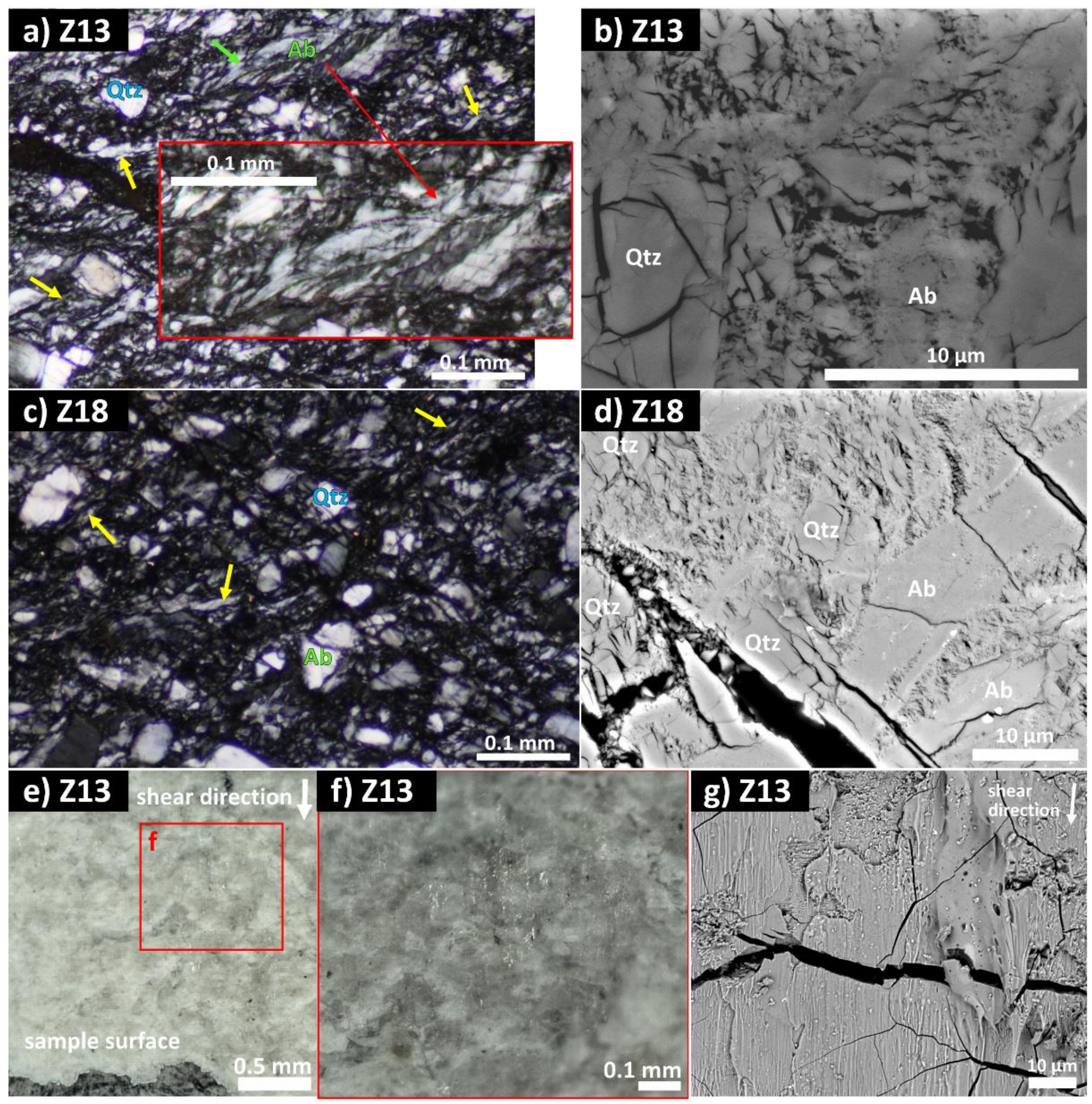

Figure 5. Photomicrographs of samples Z13 (390 °C & 344 MPa; a, b, e, f, and g) and Z18 (540 °C & 477 MPa; c and d). Figures 5a and 5c were taken with an optical microscope at Tohoku University under crossed polars. Figures 5b and 5d were taken with an FE-SEM at Tohoku University. Figures 5e and 5f were taken with an incident light microscope at GSJ-Lab of AIST, and Figure 5g was taken with a W-tip scanning electron microscope at GSJ-Lab of AIST. (a and c) Relatively coarse grains are found between fine grains, and some fine grains show elongations (yellow arrows). The insert in Figure 5a is a magnified view of an elongated albite grain. (b and d) Quartz grains exhibit abundant intragranular cracks, while albite grains have nanometer-sized pores. Figure 5b is a brightened image of the original one produced by adding a grayscale of 40. (e and f) The surface of the piston-sample boundary exhibits milky-white substances aligned sub-parallel to the shear direction. The area shown in Figure 5f is indicated by a red rectangle in Figure 5e. (g) Nanometer-wide fibers on the surface of the piston-sample boundary. The shear sense in Figures 5a, and 5c, are approximately dextral. The shear directions in Figures 5b, 5d, 5e, 5f, and 5g are shown with white arrows. Qtz: quartz, Ab: albite.

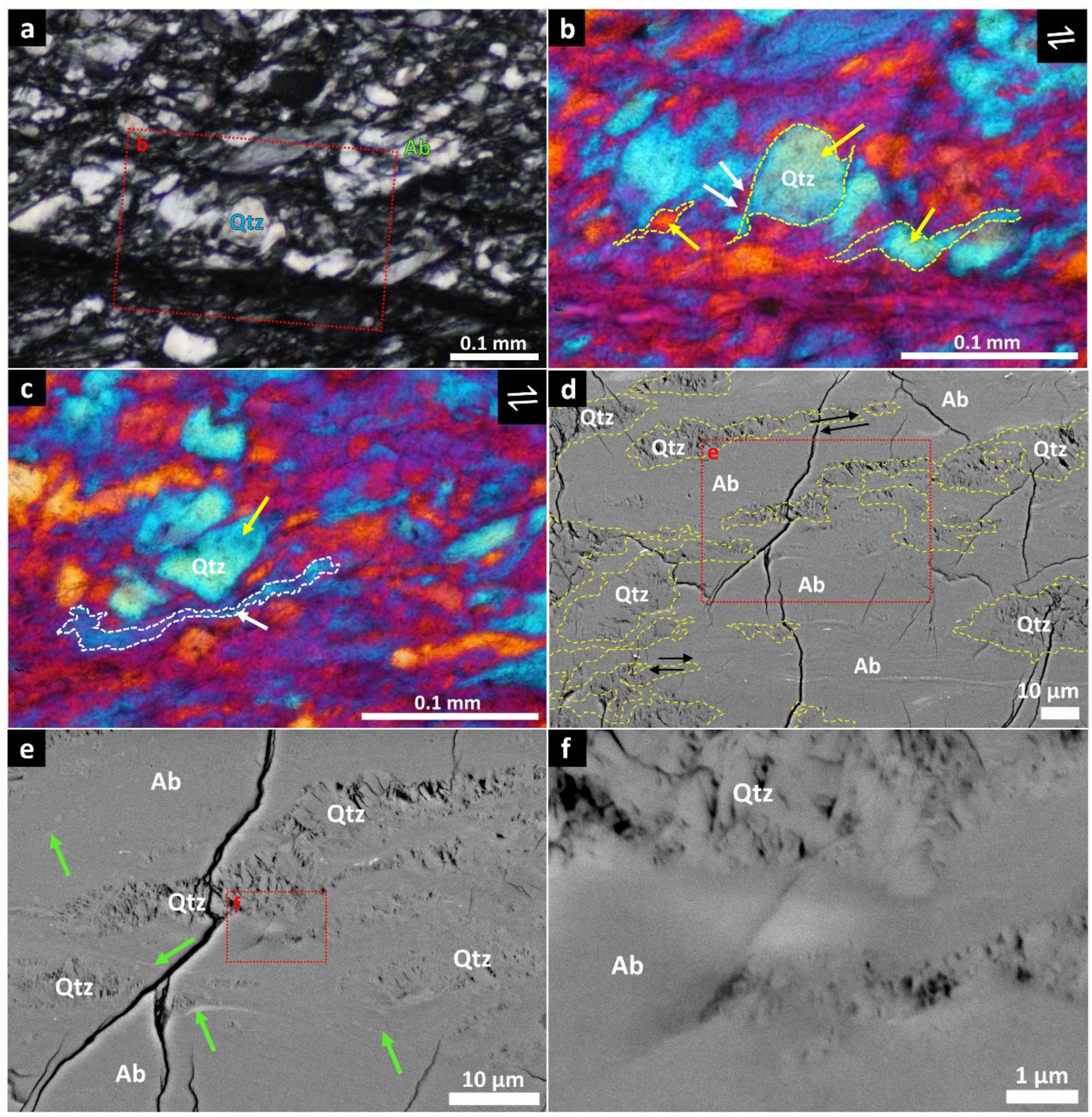

Figure 6. Microstructures of sample Z24 (720 °C & 750 MPa). Figures 6a, 6b, and 6c were taken with an optical microscope under crossed polars. For Figures 6b and 6c, a test plate ($\lambda$ = 530 nm) was inserted. Figures 6d, 6e, and 6f are BSE images. (a) The area of localized shear deformation. The area magnified in Figure 6b is shown with a red rectangle. (b) The outlines of the porphyroclasts (yellow lines) show cores (yellow arrows) having two tails. Rounded fine grains (white arrows) surround a large core. (c) Angular grain with a sharp tip (yellow arrow) and an elongated grain (the area surrounded by white line) having a bent part (white arrow). (d) Based on Al and Na maps, possible boundaries of quartz and albite domains were drawn with yellow lines. The area magnified in Figure 6e is shown with a red rectangle. The pairs of black arrows denote the local shear directions estimated from the elongated structures. (e) Elongated bright structures are shown with green arrows. The area magnified in Figure 6f is shown with a red rectangle. (f) Image taken at a magnification of ×10000. Pores are not visible in the albite domain. All images except for Figures 6b and 6c are the dextral sense of shear, and the shear direction in Figures 6b and 6c is shown with white arrows in the upper right corner. Qtz: quartz, Ab: albite.

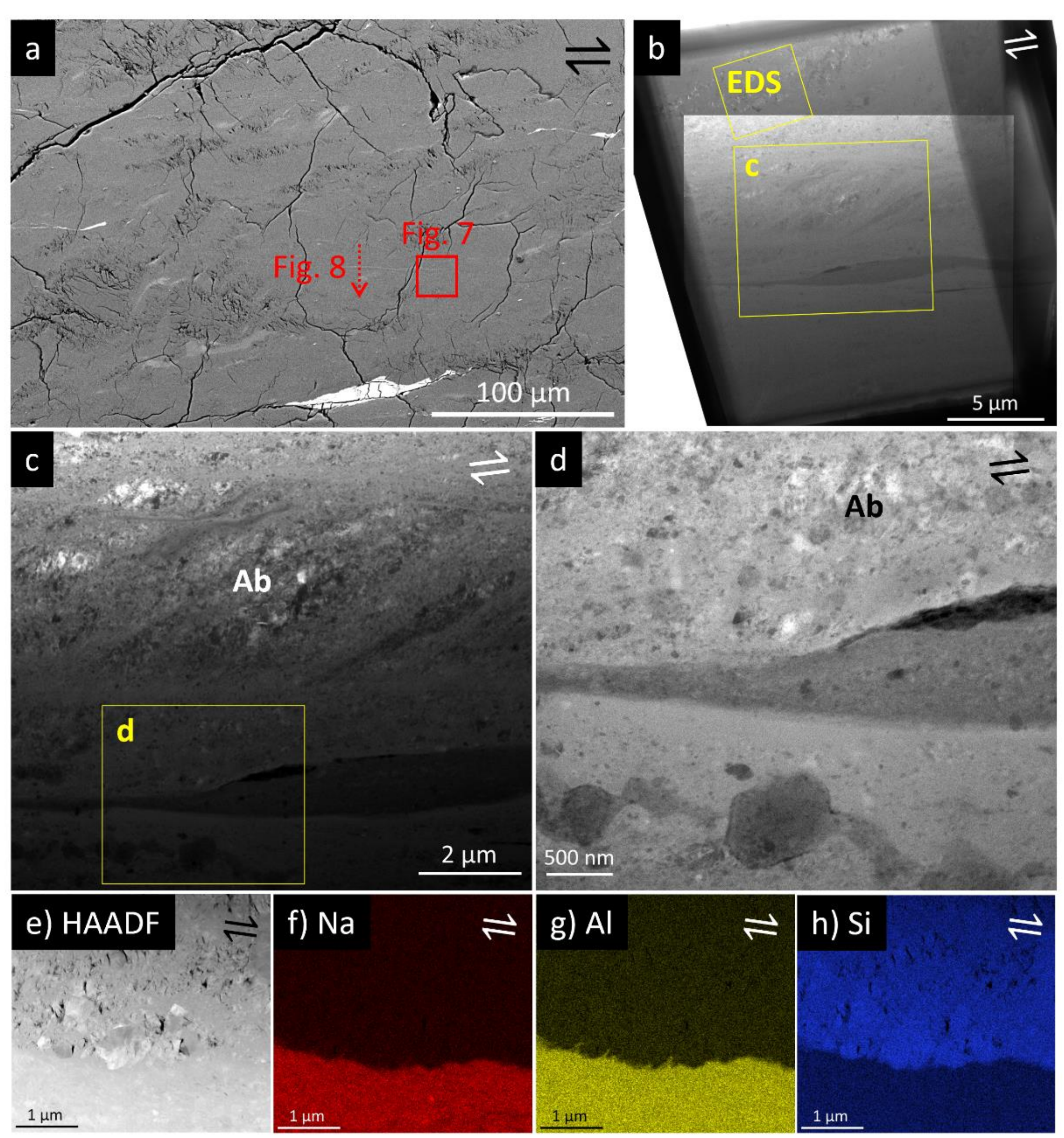

Fig. 7 Furukawa et al. (2025)

Figure 7. Micro- and nano-structures of the low-viscosity looking material of sample Z24 (720 °C & 750 MPa). (a) SEM- BSE image that shows the locations of the two FIB-sections shown in Figures 7 and 8. (b) TEM image of the overview of the FIB-section that was cut parallel to the half-cut plane. The brightness was adjusted to clarify the structure. Rectangles show the locations of Figure 7c and EDS maps (Figures 7e-h). (c and d) Enlarged TEM images. Figure 7c shows that the albite domains

are extended and consist of nano-grains (< 100 nm). The rectangle shows the location of Figure 7d, which shows nano-grains. (e-f) HAADF image (Figure 7e) and EDS element maps obtained there. Na (Figure 7f), Al (Figure 7g), and Si (Figure 7h) distributions exhibit a sharp boundary between quartz and albite domains. Shear direction is shown in each image. Ab: albite.

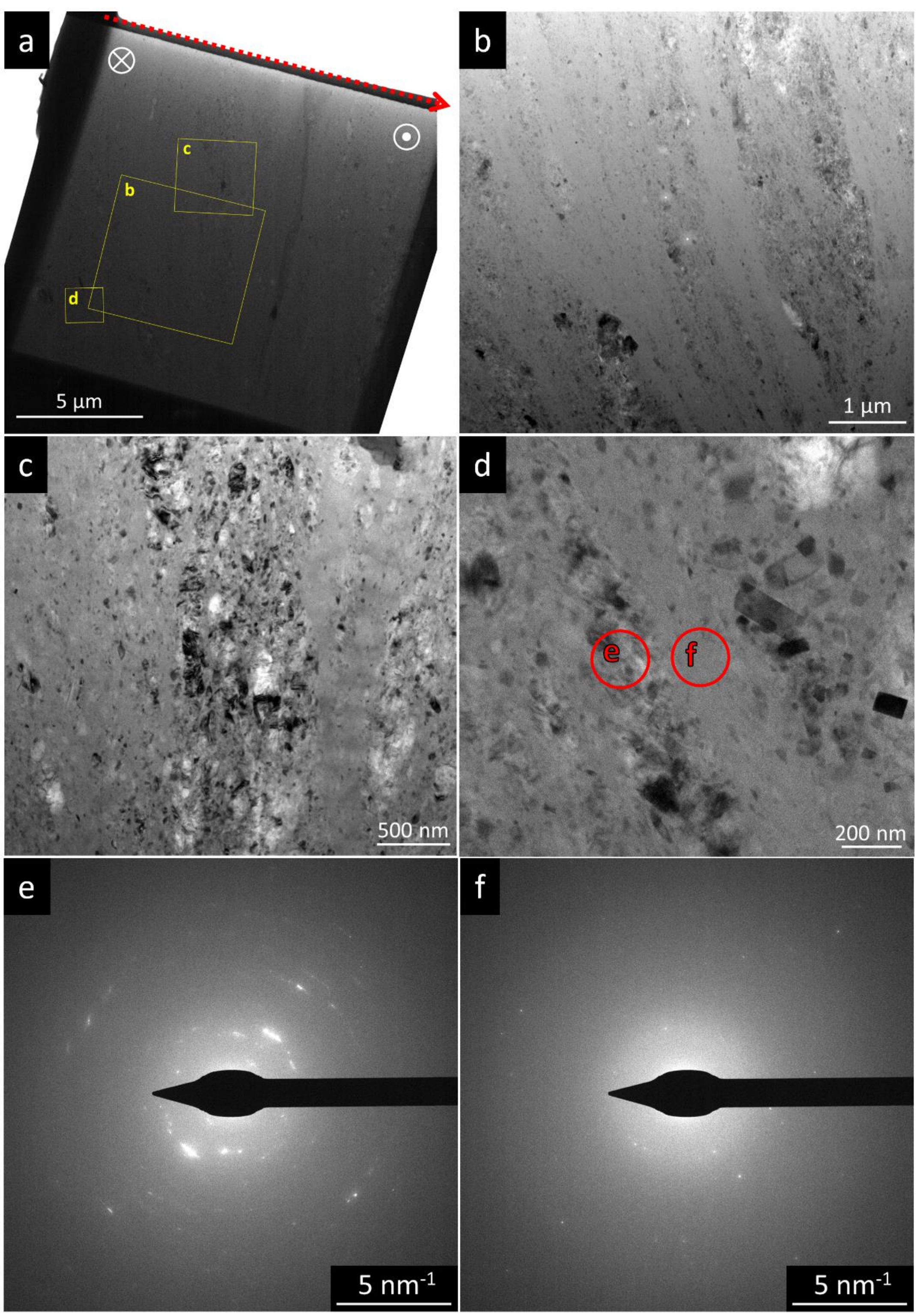

Fig. 8 Furukawa et al. (2025)

Figure 8. Micro- and nano-structures in sample Z24 (720 °C & 750 MPa) of the area shown with the red arrow in Figure 7a. (a) TEM image of the overview of the FIB-section. Brightness was adjusted to secure clarity of the image. Cross and dot in the circle denote shear direction (cross: from front to back, dot: from back to front). Rectangles show the areas of Figures 8b-8d. (b) Difference in color on the image shows extended domains, suggesting flow. The domains include nano-grains. (c and d) Magnified views of the areas shown in Figure 8a. Red circles in Figure 8d denote the areas where the diffraction patterns of Figures 8e & 8f were taken. A relatively weak diffraction area (Figure 8f) exists next to a relatively strong diffraction area (Figure 8e).

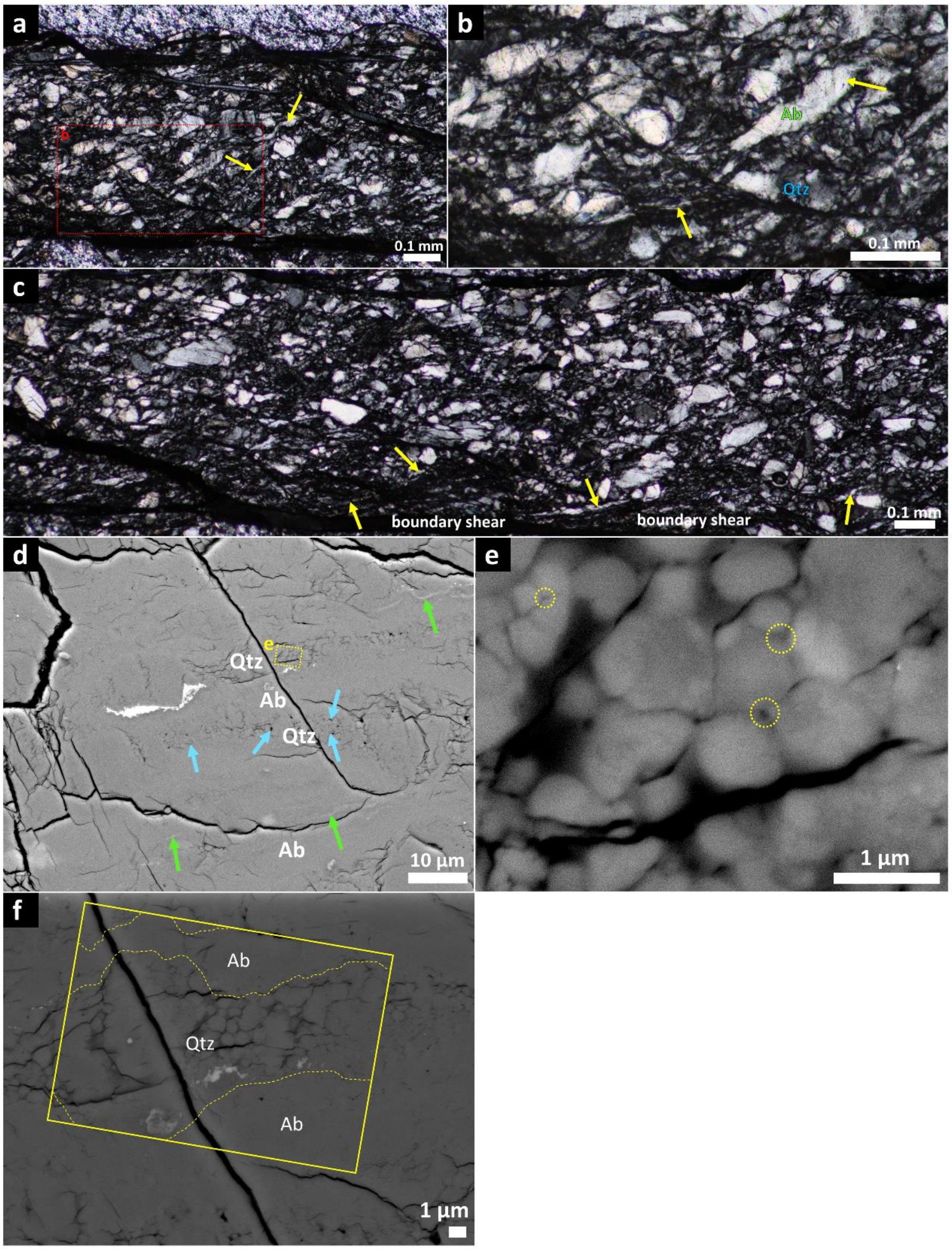

Fig. 9 Furukawa et al. (2025)

Figure 9. Microstructures of sample Z30 (900 °C & 870 MPa). Figures 9a, 9b, and 9c were taken with an optical microscope under crossed polars. Figures 9d, 9e, and 9f are BSE images. (a and b) Rounded fine grains surround coarse grains. Yellow arrows denote elongated grains. The area magnified in Figure 9b is shown with a red rectangle in Figure 9a. (c) Photomicrographs stitched to cover the entire boundary shear zones. Elongated grains (yellow arrows) are observed in the boundary shear zones. (d) The quartz domains consist of polygonal grains, while the albite domains have few pores and apparently homogeneous. Elongated bright structures are shown with green arrows, and pores between the quartz polygonal grains are shown with blue arrows. The area magnified in Figure 9e is shown with a yellow rectangle. (e) BSE image of the area composed of the quartz polygonal grains, as denoted in Figure 9d. Yellow circles denote triple junctions with 120° dihedral angles. (f) BSE image of the area where the element mapping was captured. The rectangle shows the location of the element map, and the yellow dotted lines denote mineral phase boundaries based on the Na map. Figures 9a and 9c were rotated by 1° and Figure 9b was rotated by 9° from original images to align the shear direction to the right on the image. Figure 9d is the dextral sense of shear. Qtz: quartz, Ab: albite.

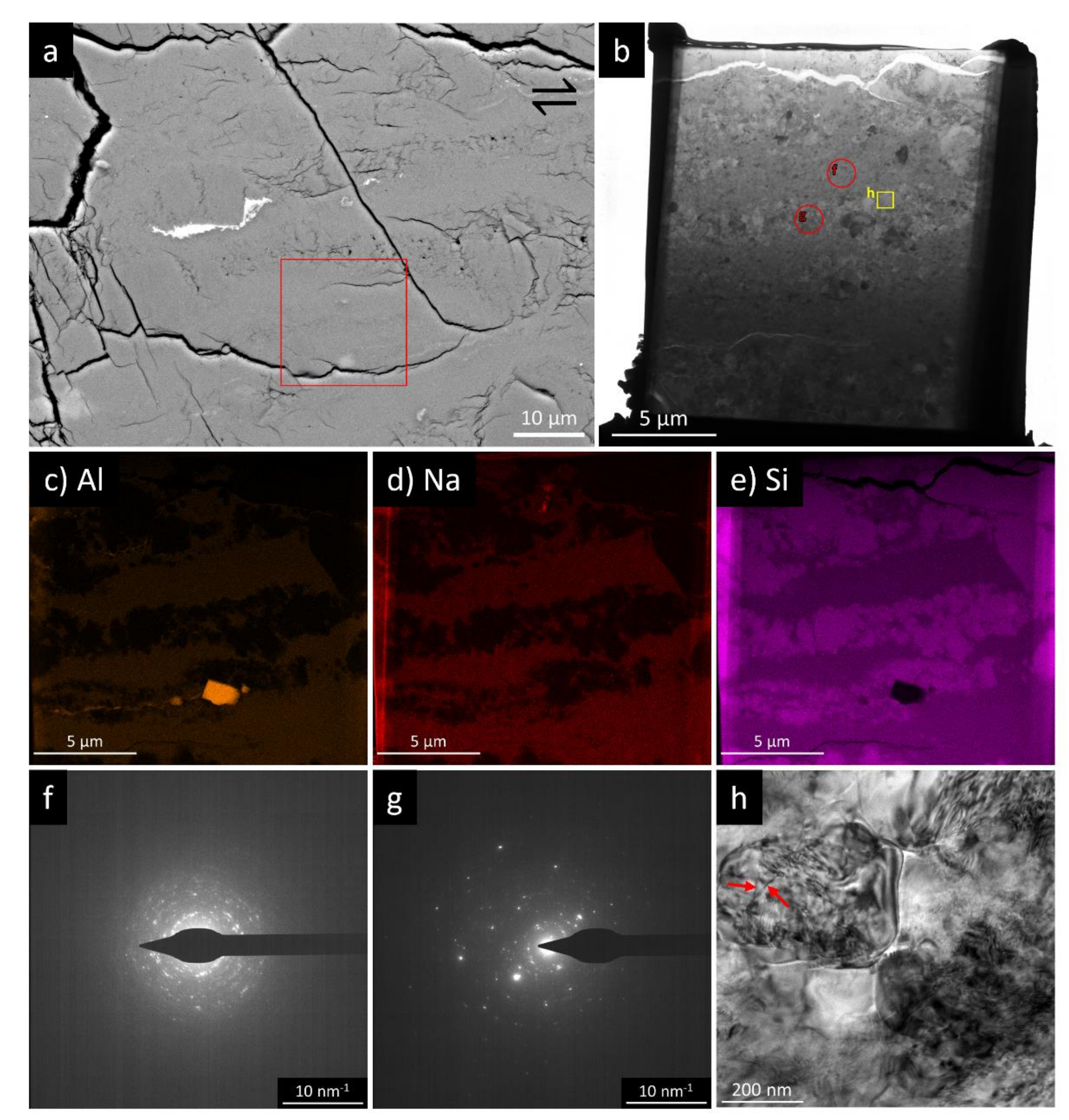

Figure 10. Micro- and nano-structures of sample Z30 (900 °C & 870 MPa). (a) SEM BSE image around area that the FIB-section was made, shown with a rectangle. The pair of arrows show the shear direction. (b) TEM image of the overview of the FIB-section. Brightness was increased to ensure a clearer view. The circles denote where the diffraction patterns of Figures. 10f and 10g were taken, and the rectangle is the area of Figure 10h. (c-e) EDS maps of Al (Figure 10c), Na (Figure 10d), and Si

(Figure 10e). (f and g) Diffraction patterns at albite domain (Figure 10f) and quartz domain (Figure 10g). Both patterns suggest polycrystalline. (h) Bright field TEM image at the area shown in Figure 10b. Crystal defects are observed in a grain (red arrows).

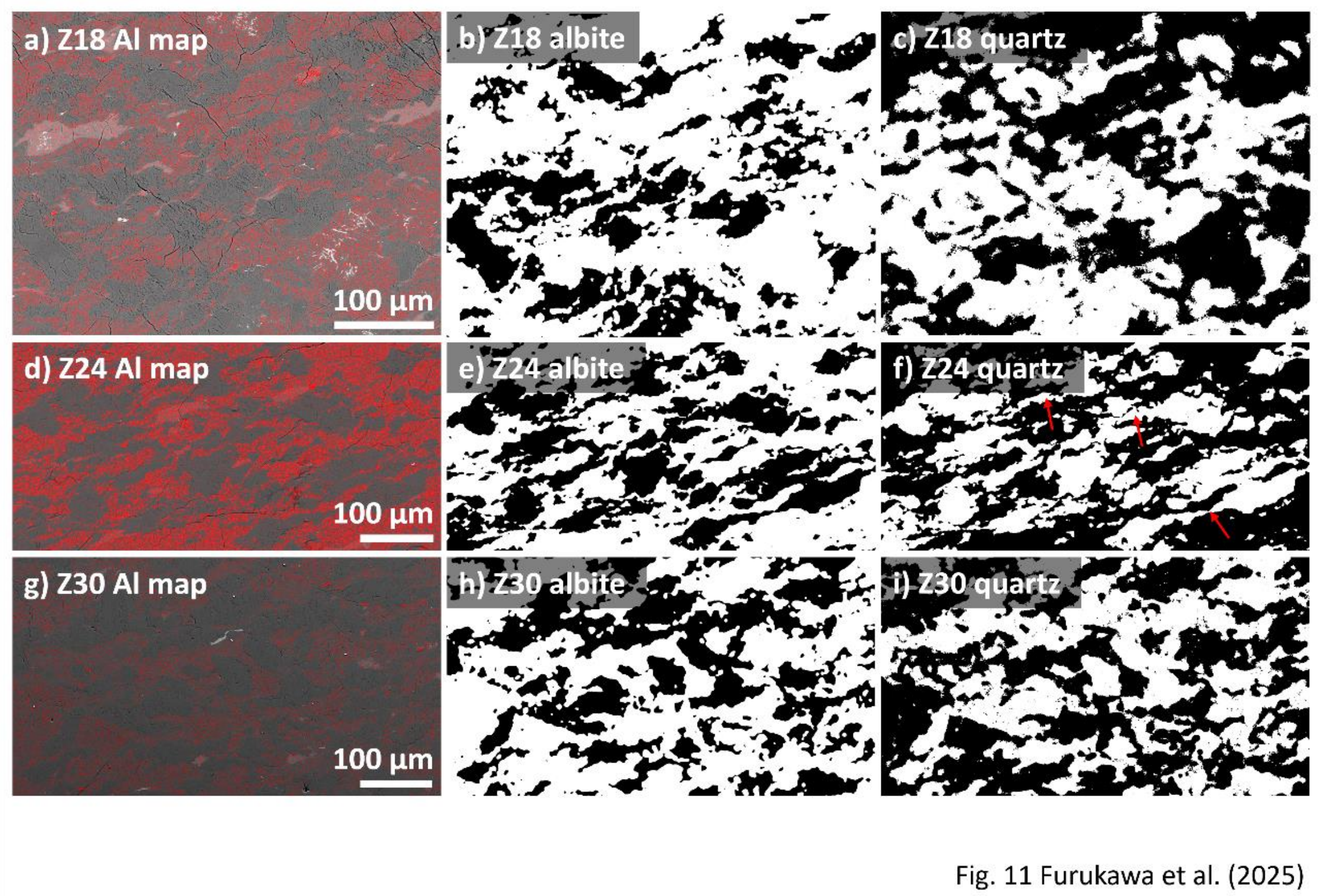

Figure 11. Binary images used for the calculation of the ACF, and original aluminum maps covering the same areas as the corresponding binary image. As for the binary images, the extracted phases are shown in white. The shear direction of all images is dextral. (a, d, and g) Cropped and rotated original Al maps to show exactly the same areas as the binary images on the right side of the same column in Figure 11. The mapping plots are shown in red. (b and c) Binary images used for the ACF calculation of albite distribution (Figure 11b) and quartz distribution (Figure 11c) in sample Z18. (e and f) Binary images used for the ACF calculation of albite distribution (Figure 11e) and quartz distribution (Figure 11f) in sample Z24. The red arrows in Figure 11f denote elongated domains connecting the coarse grains. (h and i) Binary images used for the ACF calculation of albite distribution (Figure 11h) and quartz distribution (Figure 11i) in sample Z30.

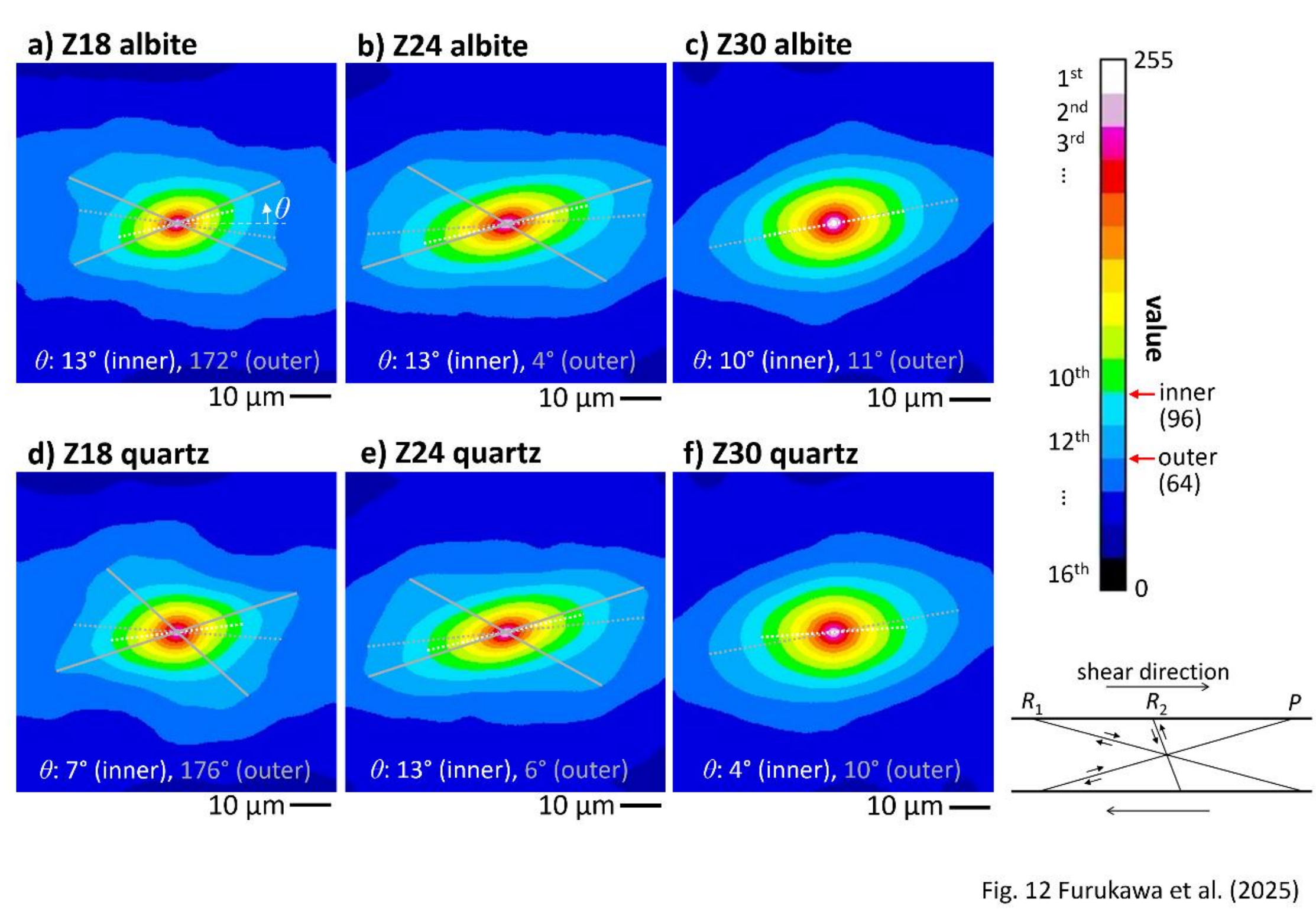

Figure 12. Contour maps showing the results of the calculation of the autocorrelation function (ACF). The white dotted lines and gray dotted lines denote the directions of the major axis of ellipses fit to the inner contours and the outer contours, respectively. The direction $\theta$ is measured in a counterclockwise direction from the horizontal right, as suggested in Figure 12a. The gray solid lines in Figures 12a, 12b, 12d, and 12e denote visually determined directions of the bulges of the outer contours. The results of (a) albite map for sample Z18, (b) albite map for sample Z24, (c) albite map for sample Z30, (d) quartz map for sample Z18, (e) quartz map for sample Z24, and (f) quartz map for sample Z30. The values of the calculation results were divided into 16 classes with different colors, as denoted on the upper right. Red arrows next to the color bar denote the values where the inner and outer contours were defined to compare the aspect ratio and the major axis direction of the fit ellipses. For all contour maps, the length of each side is 256 pixels, and the shear sense is dextral. The inserted

figure on the lower right shows a schematic modified from Logan et al. (1979), showing the directions of the $R_1$ shear, $R_2$ shear, and $P$ shear using the terminology in Logan et al. (1979).

Table 2. Summary of aspect ratio and major axis direction of fit ellipses in ACF analysis

| Sample name | Phase mineral | Inner contour | | Outer contour | |
|---|---|---|---|---|---|
| | | Aspect ratio | Major axis direction $\theta$ (°) | Aspect ratio | Major axis direction $\theta$ (°) |
| Z18 | albite | 1.74 | 13 | 1.62 | 172 |
| Z18 | quartz | 1.72 | 7 | 1.60 | 176 |
| Z24 | albite | 2.37 | 13 | 2.21 | 4 |
| Z24 | quartz | 2.34 | 13 | 2.25 | 6 |
| Z30 | albite | 1.66 | 10 | 1.88 | 11 |
| Z30 | quartz | 1.58 | 4 | 1.86 | 10 |

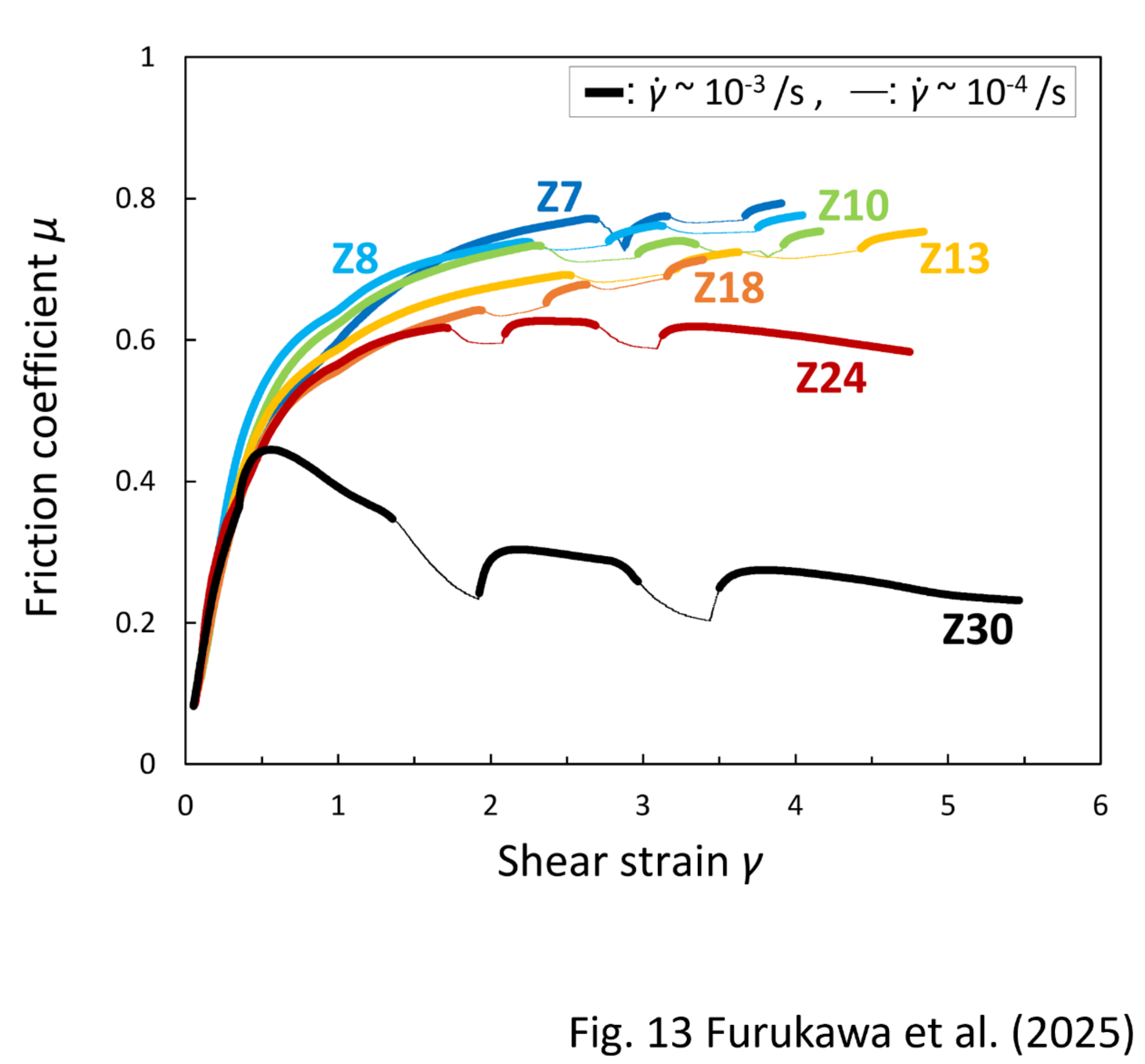

Figure 13. Relationship between shear strain $\gamma$ and friction coefficient $\mu$. The friction coefficient values were calculated using the shear stress and the average value of the confining pressure. Data were smoothed with a moving average of 30, in the same way as for the shear stress-shear strain curves (Fig. 2). Bold lines: shear strain rate of ~ $10^{-3}$ /s, thin lines: shear strain rate of ~ $10^{-4}$ /s. The lines in blue: Z7, light blue: Z8, light green: Z10, yellow: Z13, orange: Z18, red: Z24, and black: Z30.

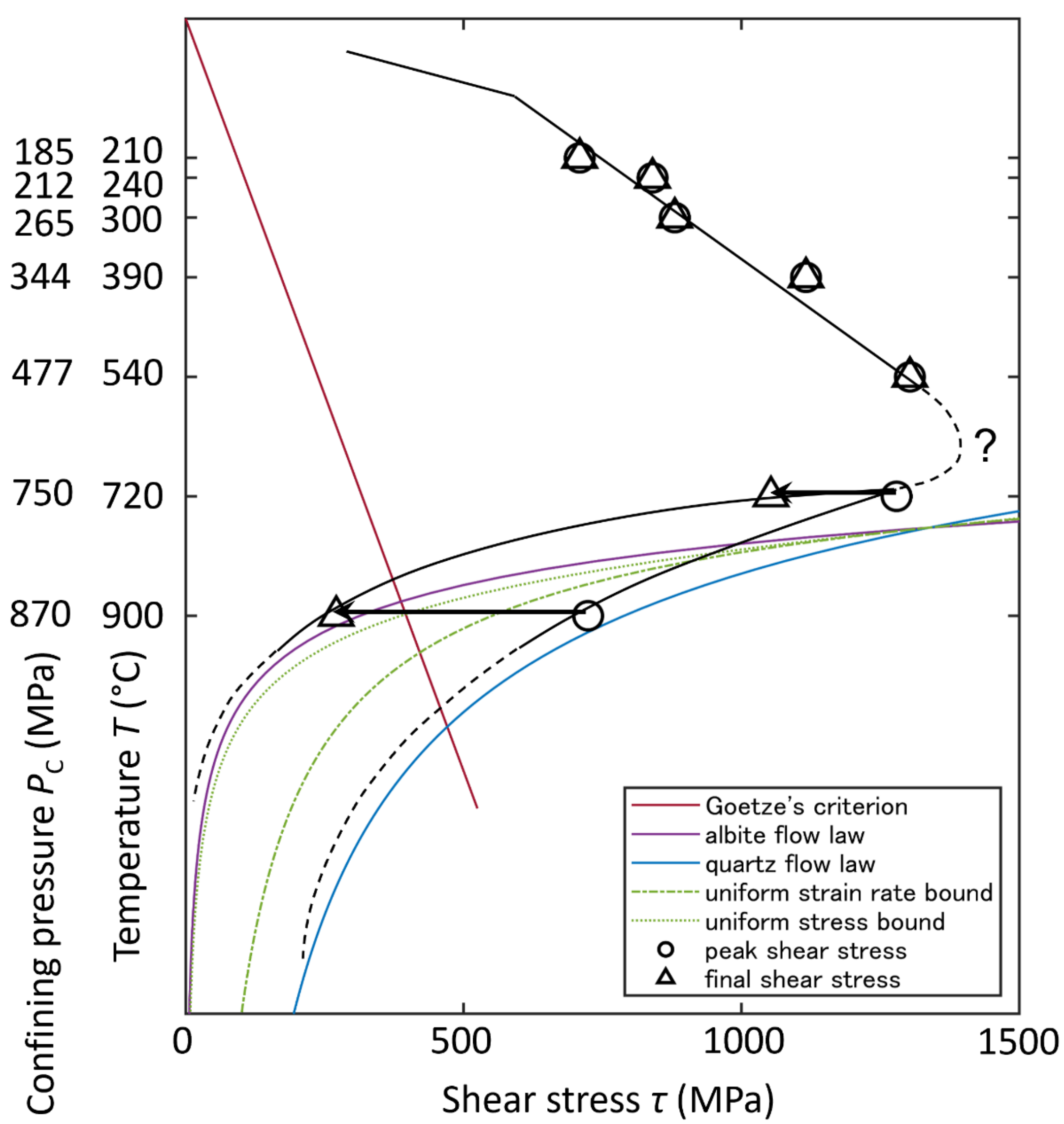

Fig. 14 Furukawa et al. (2025)

Figure 14. Relationships between the shear stress $\tau$ and the simulated depth. The experimental data of the peak shear stress are plotted with the open circle, and those of the final shear stress are plotted with the open triangles. Black arrows connect the peak and final shear stresses of the same experiments (experiment Z24 or Z30), indicating the extent of strain weakening. The black lines show an estimation of the strength profile of the upper crust, by visually connecting the experimental data. The portions

shown with the black dotted lines indicate that the stress values there are relatively uncertain, as the experiment conditions corresponding to depths there are lacking in this study. Red line shows the Goetze's criterion (half of the theoretical confining pressure). Purple and blue curves show mineral flow laws of albite and quartz at the faster strain rate ($1.4 \times 10^{-3}$ /s), respectively. The flow law parameters are derived from Offerhaus et al. (2001) in Rybacki & Dresen (2004) for albite, and revised Luan & Paterson (1992) in Fukuda & Shimizu (2017) for quartz. Green dotted lines denote two end members of the flow law of the 50-50 quartz-albite mixtures, which are the uniform stress bound and the uniform strain rate bound, calculated using equations in Tullis et al. (1991).

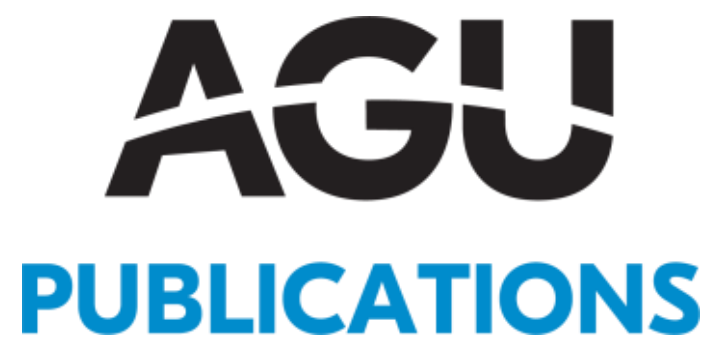

# The brittle-plastic transition in quartz-albite mixtures: New insights from shear deformation experiments at mid-to-lower crustal depth conditions

Miho Furukawa[1], Berend A. Verberne[2], Sando Sawa[1], Hiroyuki Nagahama[1], Miki Takahashi[2], Oliver Plümper[3,4], Jun Muto[1]

[1] Department of Earth Science, Graduate School of Science, Tohoku University, Sendai, Japan

[2] Geological Survey of Japan, The National Institute of Advanced Industrial Science and Technology, Tsukuba, Japan

[3] Department of Earth Sciences, Utrecht University, Utrecht, The Netherlands

[4] Faculty of Geosciences and MARUM - Center for Marine Environmental Sciences, University of Bremen, Germany

## Contents of this file

## Introduction

The supporting information includes texts explaining the experimental method (Text S1), the calculation of the initial thickness of the sample layer (Text S2), the details of the microscopy (Text S3), another procedure of phase segregation using element maps to confirm the validity of the method in this study (Text S4), a detailed explanation for the procedure of segregating mineral phases (Text S5), the procedure of obtaining ACF contour maps (Text S6), and the way of removing the data of the period during the axial loading stopped in experiment Z30 (Text S7). It also includes the grain size distributions of the starting material (Fig. S1), the flow chart showing the general procedure of the image analyses (Fig. S2), comparisons of the maps of mineral phase distributions obtained from the two different analysis procedures (Fig. S3), the phase plotting maps and the trained regions used in the machine learning performed for the phase segmentation (Fig. S4), a summary showing the process of the data removal for experiment Z30 (Fig. S5), the comparisons of the EDS results obtained by SEM and TEM analyses of samples Z24 and Z30 (Fig. S6), and the calculated quartz grain growth during the quench of experiment Z30 (Fig. S7).

## Text S1.

The experimental assembly and procedures are described by Kido et al. (2016). Axial load was measured using a load cell located above the loading column. A hydraulic pressure ram controlled confining pressure. Displacement was measured with a displacement transducer, located close to the load cell.

Inner salt sleeves were made from potassium iodide (KI) for the experiments conducted under temperatures lower than 700 ° C (experiments Z7, Z8, Z10, Z13, and Z18). On the other hand, experiments under a temperature higher than 700 ° C (experiments Z24 and Z30) used sodium chloride (NaCl) inner sleeves to avoid melting because the melting point of KI is 680 ° C at ambient pressure (Sato et al., 1987). Outer salt sleeves were made from NaCl, with an S-type thermocouple inserted. The sample was placed evenly between a pair of alumina pistons cut at 45° with respect to the direction of the loading axis. Using a diamond blade, the pre-cut surfaces were manually grooved to achieve a firm grip between the sample powder and pistons. The sample and pistons were encapsulated in an annealed Ag (experiments conducted at temperatures lower than 800 ° C) or Pt jacket (experiments conducted at temperatures higher than 800 ° C) of 0.1 mm thickness. Both ends of the jacket were mechanically sealed with Ag or Pt disks. The sealed jacket was then put in an outer Ni jacket (Fig. 1c).

Temperature ($T$) and confining pressure ($P_c$) were sequentially increased until desired experimental conditions were achieved. At an intermediate temperature (65 ° C in Z7, 50 ° C in Z8, 150 ° C in Z10, Z13 and Z18, 300 ° C in Z24, and Z30), the axial loading piston was lowered until it touched the upper alumina piston, which is called a "cold hit." The cold hit is used to (i) search for the approximate displacement where the loading piston hits the alumina piston (i.e., the "hit point"), and (ii) to pre-compact the sample powder prior to imposing shear deformation. After the hit point determination, the piston was retracted by ~ 1 mm, followed by an increase in $T$ and $P_c$ to the desired conditions. Once reaching the desired experimental conditions, the axial loading piston was lowered, imposing shear deformation on the sample layer. After the experiments, samples were quenched within a time window of ~ 1 minute to 50 ° C for experiments Z7, Z8, Z10, and Z13, to 100 ° C for Z18. For experiment Z30, the quench to 300 ° C took ~ 20 seconds, and for experiment Z24, the time taken for the quench was not measured because the thermocouple was broken while increasing the confining pressure and the temperature prior to the experiment and thus we manually controlled the temperature by changing the output value of the electric current. In experiments Z7, Z8, Z10, Z13, and Z18, the piston was retracted simultaneously with the quench, while in experiments Z24 and Z30, it was retracted after the quench. Despite such differences, note that in both cases, the onset of quench decreased the confining pressure and the axial load. The quench was followed by further reduction of temperature and pressure to room temperature and atmospheric pressure conditions in all the experiments, and the assembly was dismounted.

Text S2.

The initial thickness of the sample layer was assumed by the following calculations. Using the values in Robie et al. (1967), the density of quartz ($\alpha$-Quartz) is 2.648 g/cm$^3$, and that of albite (Low Albite) is 2.62 g/cm$^3$. Since each experiment uses 0.05 g of quartz powder and albite powder, the total volume of the sample is ~ 38 mm$^3$. The area of the saw-cut surface of the shear piston is ~ 61 mm$^2$, assuming that the surface is flat since the amount of sample stuck in the grooves is negligibly small compared to the volume of the bulk sample as suggested by, e.g., Fig. 1d. Therefore, the initial thickness of the sample layer is calculated as 0.62 mm if the porosity is 0%. Meanwhile, if the porosity is 10% and 50%, the initial thickness is calculated as 0.69 mm and 1.2 mm, respectively. In fact, the porosity of sample Z10H is estimated to be ~ 30% using an area in Fig. 1d, assuming that pores are homogeneously distributed in the sample layer. When the porosity is 30%, the initial thickness of the sample layer is given as 0.88 mm. We hence presumed that the porosity in the other samples, which were deformed at various conditions including lower temperature and confining pressure or higher temperature and confining pressure than Z10H, would fall between the range of 10-50% at the beginning of the experiment. Therefore, we assumed that the initial thickness of the sample layer is 1 mm.

Text S3.

The recovered samples were made into thin sections and resin mounts. The samples were polished using abrasive powder (#400, 800, 2000, 3000, 4000, and 6000), diamond papers (9 μm, 2 μm, and 0.5 μm), and diamond paste (0.25 μm). Samples Z8, Z10, Z13, and Z18 were further polised using colloidal silica. For samples Z7, Z8, Z10, Z13, and Z18, loose fragments were retrieved from one of the halves to perform observations of the surface of the sample layer contacting to the shear piston. They were emplaced on an SEM stub with a carbon sticker, using carbon glue and Pt deposition to facilitate conduction.

Inter- and intra-grain microstructures were investigated in transmitted light using a polarizing light microscope (Nikon; ECLIPSE C*i* POL). Objective lenses were 4x (numerical aperture (NA): 0.10), 10x (NA: 0.25), 20x (NA: 0.40), or 40x (NA: 0.65). A field emission scanning electron microscope (SEM; JEOL JSM-7001F) installed at Tohoku University was used for imaging, equipped with an energy-dispersive spectrometer (EDS; INCA, and Aztec) for semi-quantitative element compositional analyses. The probe current used was 1.39-1.40 nA, and the acceleration voltage was 15.0 kV. The samples were carbon coated prior to SEM imaging using a Meiwafosis CC-40F carbon coater. To capture entire sample microstructures at high magnifications, BSE images were stitched after they were captured at lower magnifications (×80 and/or ×50) (Figs. 1d and 3). Detailed microstructures were observed at magnifications higher than ×10000. Since quartz and albite in our sample were difficult to distinguish on BSE images due to their similarity in brightness, EDS mapping was performed for aluminum (Al) and sodium (Na) to visualize the albite distributions. We also observed the surface of the piston-sample boundary of sample Z13, using a Keyence VHX-2000 digital microscope, in incident light, and a Hitachi SU3500 SEM operated in backscattered electron mode, both of which were installed at AIST.

We further performed a transmission electron microscopy (TEM) for samples Z24 (lower half) and Z30 at Utrecht University (Talos F200X). The foils were cut using a focused ion beam scanning electron microscope (FIB-SEM, Helios Nanolab G3 at Utrecht University) either parallel to or perpendicular to the half-cut surface. For the parallel-cut, we followed the method of Li et al. (2018).

Text S4.

It is often the case that BSE images and element maps are separately used to segment constituting mineral phases as well as cracks. However, in this study, we used maps where an element map of Al is overlaid on a BSE image, which enabled us to segment the mineral phases from one map. To confirm the validity of our method, we produced the mineral phase distribution maps for an area of sample Z18 following these different methods.

Figs. S3a and S3b are the results obtained from the general procedure, where we set a threshold on a BSE image and an element map, respectively. We first set thresholds on a BSE image to extract the distributions of apatite grains (threshold: 165-255), K-feldspar grains (threshold: 147-164), and cracks (threshold: 0-130). The thresholded ranges were determined by visual inspection. We also set a threshold to the Al map so that it covers the plots of the Al distributions. Since this Al map includes the pixels derived from both albite and K-feldspar, we subtracted the pixels segmented as K-feldspar grains from the Al plots and obtained a map showing the distribution of the albite mineral phase (Fig. S3a). Meanwhile, to obtain a map showing the distribution of the quartz mineral phase (Fig. S3b), we inverted black and white in the map of the Al plots, so that the pixels without the Al plots were shown in white. From this inverted image, the pixels segmented as apatite grains and cracks were subtracted, and we obtained the map of the quartz phase distribution (Fig. S3b).

Fig. S3 shows that the resulting maps produced from the two different procedures are similar in appearance for both the albite distributions (Figs. 6a and 6c) and the quartz distributions (Figs. 6b and 6d). Therefore, we segmented the mineral phases from a single map, where the Al distribution is overlaid on the BSE image, as shown in Fig. S2.

## Text S5.

Since the pixels of the locations with the Al plots are colored in red in the overlaid image, the R-values of an RGB image are larger than the G-values and B-values. Meanwhile, the values of R, G, and B of the pixels other than the Al plots are the same. By splitting the overlaid images into the grayscale images of each color value (i.e., R, G, and B), we separated the pixels of the Al plots from the others. In the image of the R-value, the pixels of the Al plots and impurities have relatively large grayscale values (Fig. S2b). Thus, we separated them by setting a threshold on these pixels and produced a binary image (Fig. S2c). Meanwhile, in the images of the B-value and the G-value, the pixels of the impurities have relatively large grayscale values, while those of the Al plots are relatively lower (Figs. S2d and S2f). By setting a threshold to the image of the B-value, we produced a binary image where the pixels of the impurities are segregated (Fig. S2e). Moreover, in the images of the B-value and the G-value, the pixels of the Al plots have generally lower values than those of the surrounding matrix. Therefore, we produced a binary image from the image of the G-value by setting a threshold to the pixels with relatively lower grayscale values (i.e., Al plots and cracks), and by inverting the colors of black and white, we obtained a binary image where the pixels of the surrounding matrix and the impurities are shown in white (Fig. S2g).

Then, we subtracted the binarized B-value image (Fig. S2e) from the binarized R-value image (Fig. S2c) and obtained a map showing the albite mineral distribution. The obtained map was further binarized, and a binary image where albite areas are shown in white. Below, we call this binary image an albite plotting map. Similarly, we obtained a map showing the quartz mineral distribution by subtracting the binarized B-value image (Fig. S2e) from the binarized and inverted G-value image (Fig. S2g). Note that some pixels belonging to K-feldspar grains and cracks could not be removed perfectly, though we consider their effects to be negligible due to their small proportion of the total. The resulting image was further binarized to obtain a binary image where quartz-derived areas are shown in white. Below, we call this binary image a quartz plotting map.

Although the actual phase distributions are continuous areas in the sample layer, the albite phase distributions are represented as a gathering of discrete plots in the albite and quartz plotting maps, reflecting the individual plots of the original element map. To fill in the pixels between the plots, we used “Trainable Weka Segmentation” (Arganda-Carreras et al., 2017), a Fiji plugin that can be used to identify whether a pixel belongs to the region of a certain phase. For an albite plotting map, we randomly selected three locations from areas that appeared to be in the albite portion and trained them as Class 1. Another three locations that did not appear to be in the albite portion were selected at random and trained as Class 2. Similarly, for a quartz plotting map, we trained three areas that appeared to be in the quartz portion as Class 1, and another three areas that did not appear to be within the quartz portion as Class 2. The trained regions for each image are shown in Fig. S4. Using the trained data, we created a classifier model by Fast Random Forest and obtained a probability map, which shows the probability that each pixel on the image belongs to Class 1. In the probability map, pixels with a probability of 1 were shown in white, pixels with a probability of 0 were shown in black, and pixels in between probability were shown in grayscale according to their value. The probability map was rotated so that the right horizontal direction became the shear direction.

Text S6.

Using the rotated probability maps in Text S5, ACF contour maps were generated in the following direction. First, the rotated probability maps were binarized to be used in the ACF analyses. To prevent the cracks from affecting the result of the analyses, which focused on phase geometry, a portion of the binarized image that did not have notable cracks was chosen and used for further calculation. Then, the ACF was calculated within each of the 256-pixel squares arranged on the image, and the calculation results were shown in grayscale maps. The resulting maps of all the regions were overlaid and averaged. To highlight the distribution of grayscale values on the maps, the values were binned into 16 classes, and their distributions are shown as contour maps where each class is color-coded.

Text S7.

We removed the data of axial stress and displacement from the period of the gear stop in experiment Z30 (Fig. S5a). We also removed data from the period just after the restart of gear rotation until the axial stress returned to the level just before the gear stop (Fig. S5a), so that the value of the axial stress smoothly increases with time (Fig. S5b). The experimental data set of Z30 was generated by connecting data just before and after the removed part (Fig. S5b).

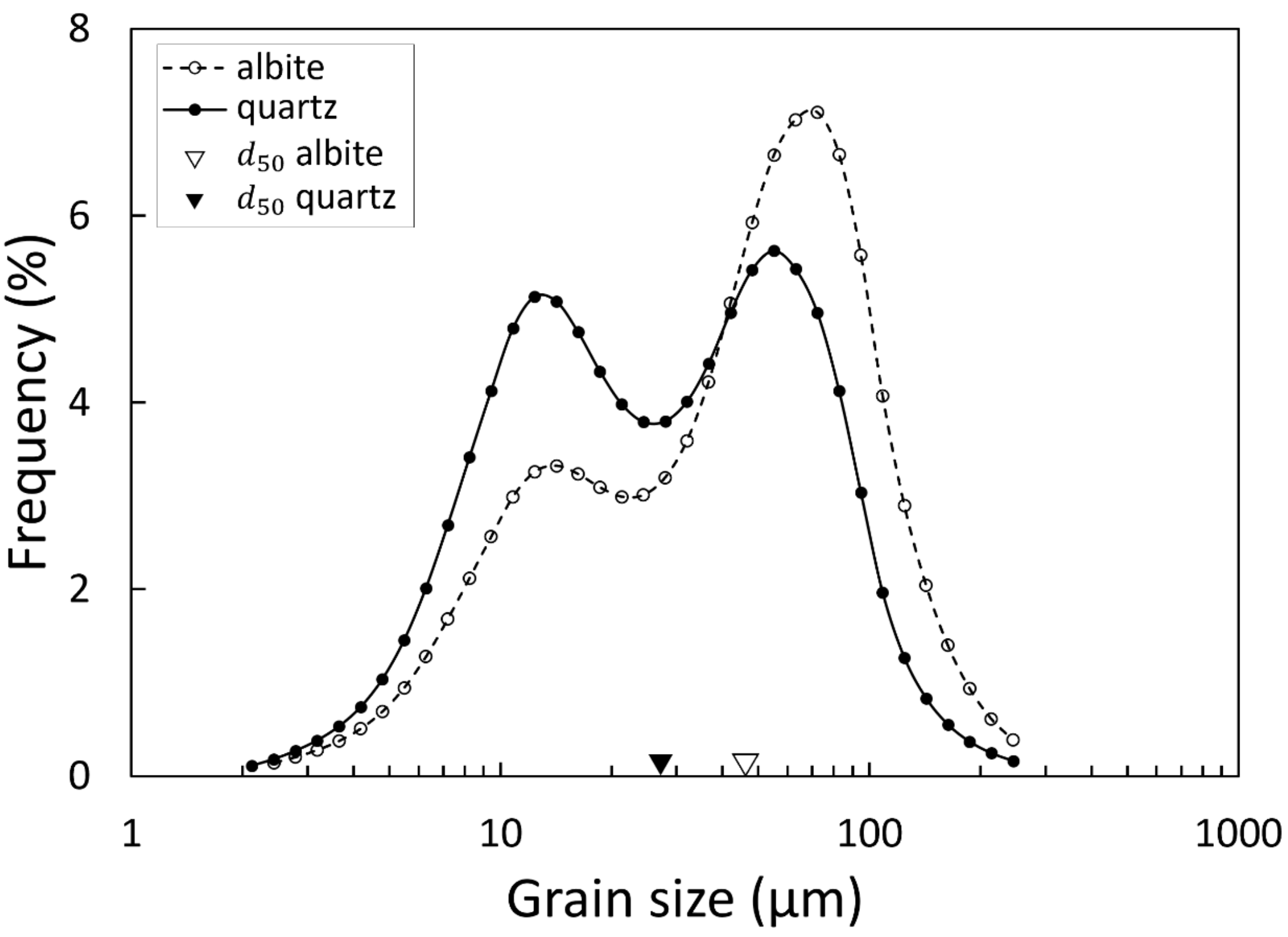

Fig. S1 Furukawa et al. (2025)

Figure S1. Grain size distributions of the starting material obtained from the laser particle size analysis. The plots of the quartz data (black circles) are connected by a smooth line shown with a solid line, while those of the albite data (open circles) are connected by a smooth line shown with a broken line. The medians of quartz and albite grain sizes ($\boldsymbol{d_{50}}$) are denoted by the black triangle and the open triangle, respectively.

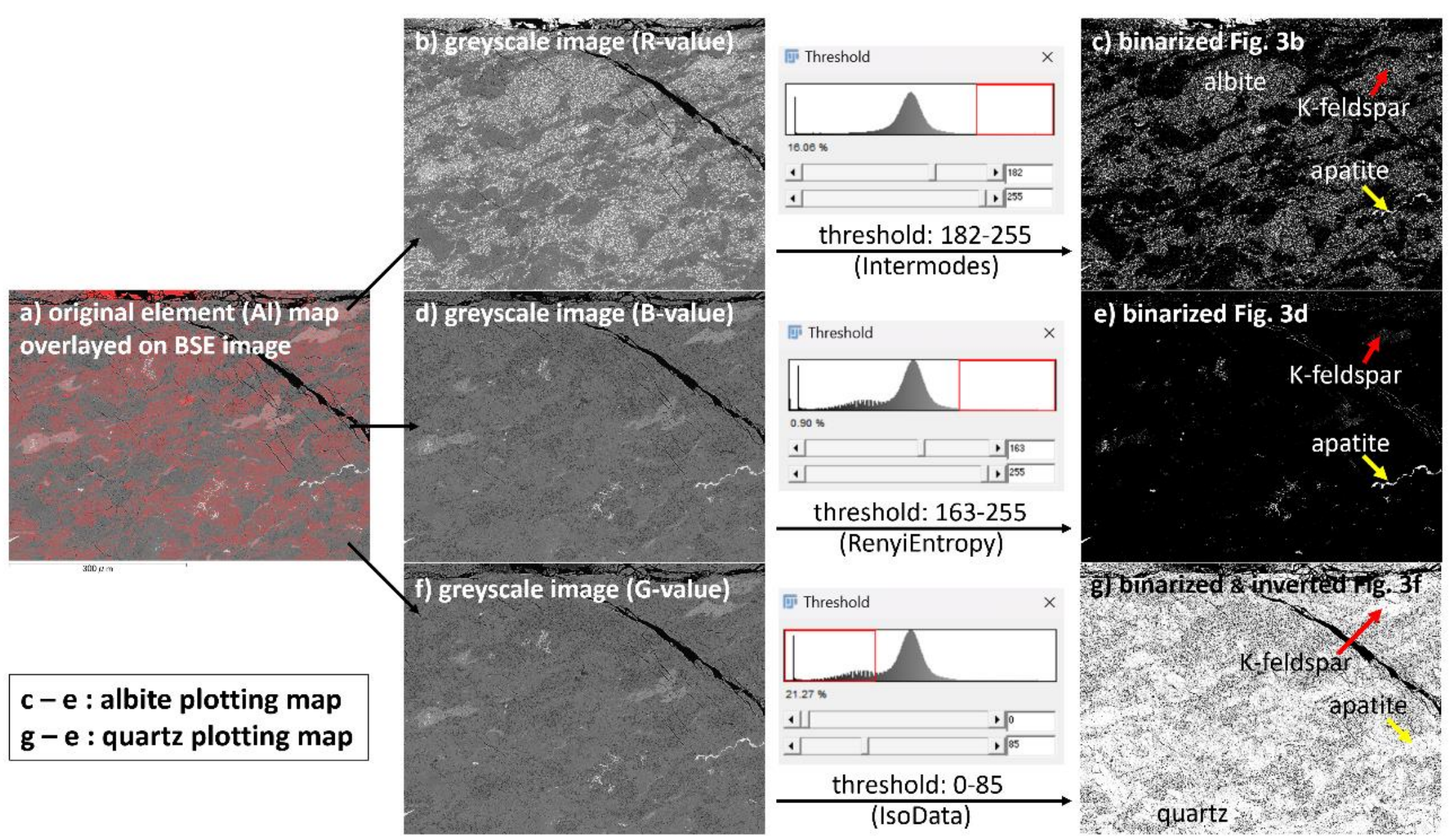

Fig. S2 Furukawa et al. (2025)

Figure S2. Flow chart showing the general procedure of the image analyses to produce albite and quartz phase maps. The images are the case of sample Z18. (a) Original element (Al) map. (b, d, and f) The original map is split into the maps of red channel values (Figure S2b), blue channel values (Figure S2d), and green channel values (Figure S2f). (c, e, and g) Binarized maps of the red channel values (Figure S2c), of the blue channel values (Figure S2e), and of the green channel values (Figure S2g). Red arrows denote a K-feldspar grain and yellow arrows denote an apatite grain. The albite plotting map is produced by binarizing the resulting image of the subtraction of Figure S2e from Figure S2c, and the quartz plotting map is given by binarizing the resulting image of the subtraction of Figure S2e from Figure S2g. Inserted figures between Figures S2b and S2c, Figures S2d and S2e, and Figures S2f and S2g show the thresholding methods used to binarize each image. The scale bar is the same for all images.

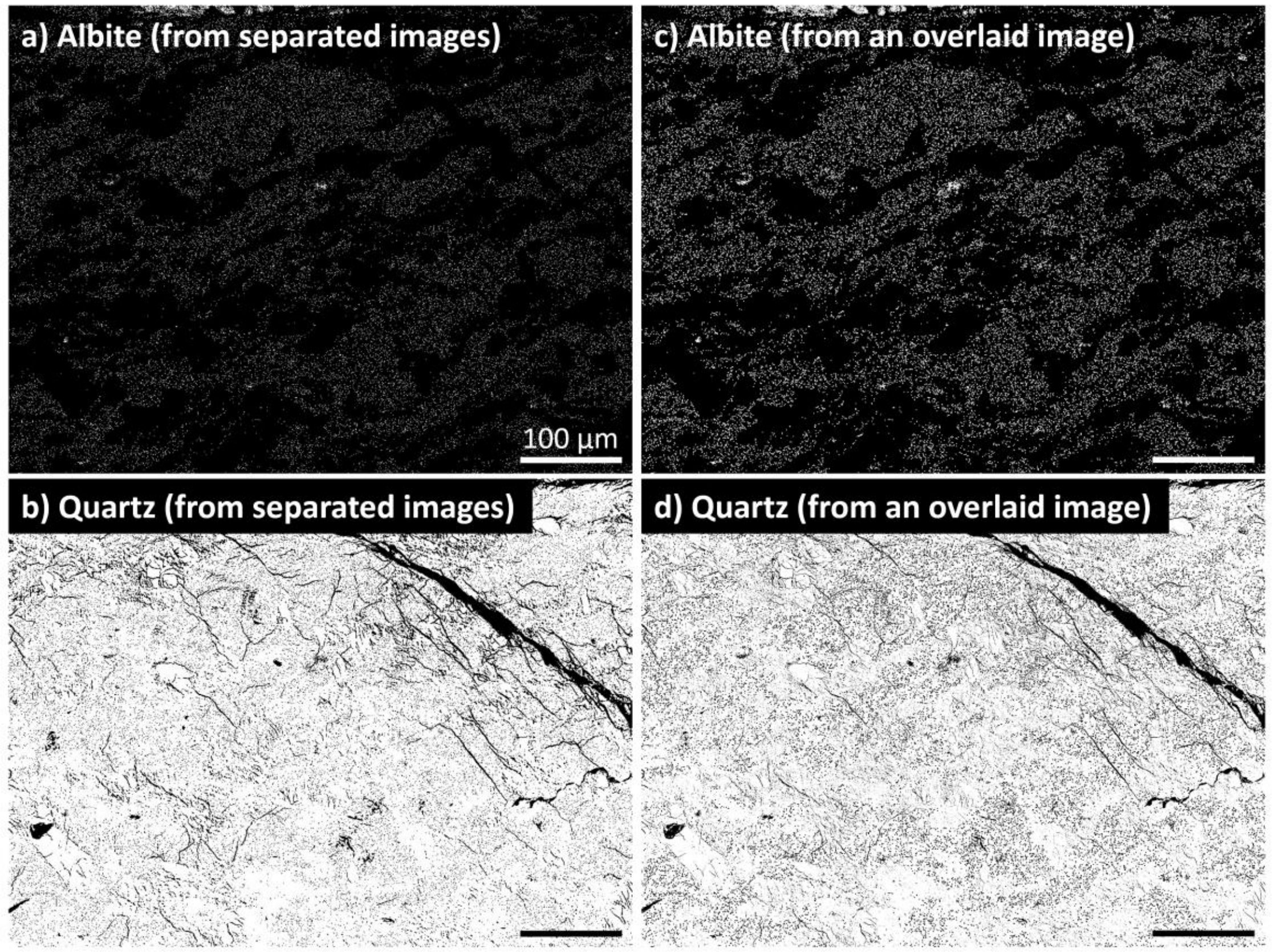

Figure S3. Comparisons between the maps of mineral phase distributions obtained from the two different analysis procedures described in Text S4. Figures S3a and S3b are obtained from the general procedures, setting thresholds to a BSE image and an element map, respectively. Meanwhile, Figures S3c and S3d are produced by the procedure in this study, using an image where the element map is overlaid on the BSE image. Figures S3a and S3c show the maps of the albite distributions, and Figures S3b and S3d are the maps of the quartz distributions. The extracted mineral phases (either albite or quartz) are shown in white.

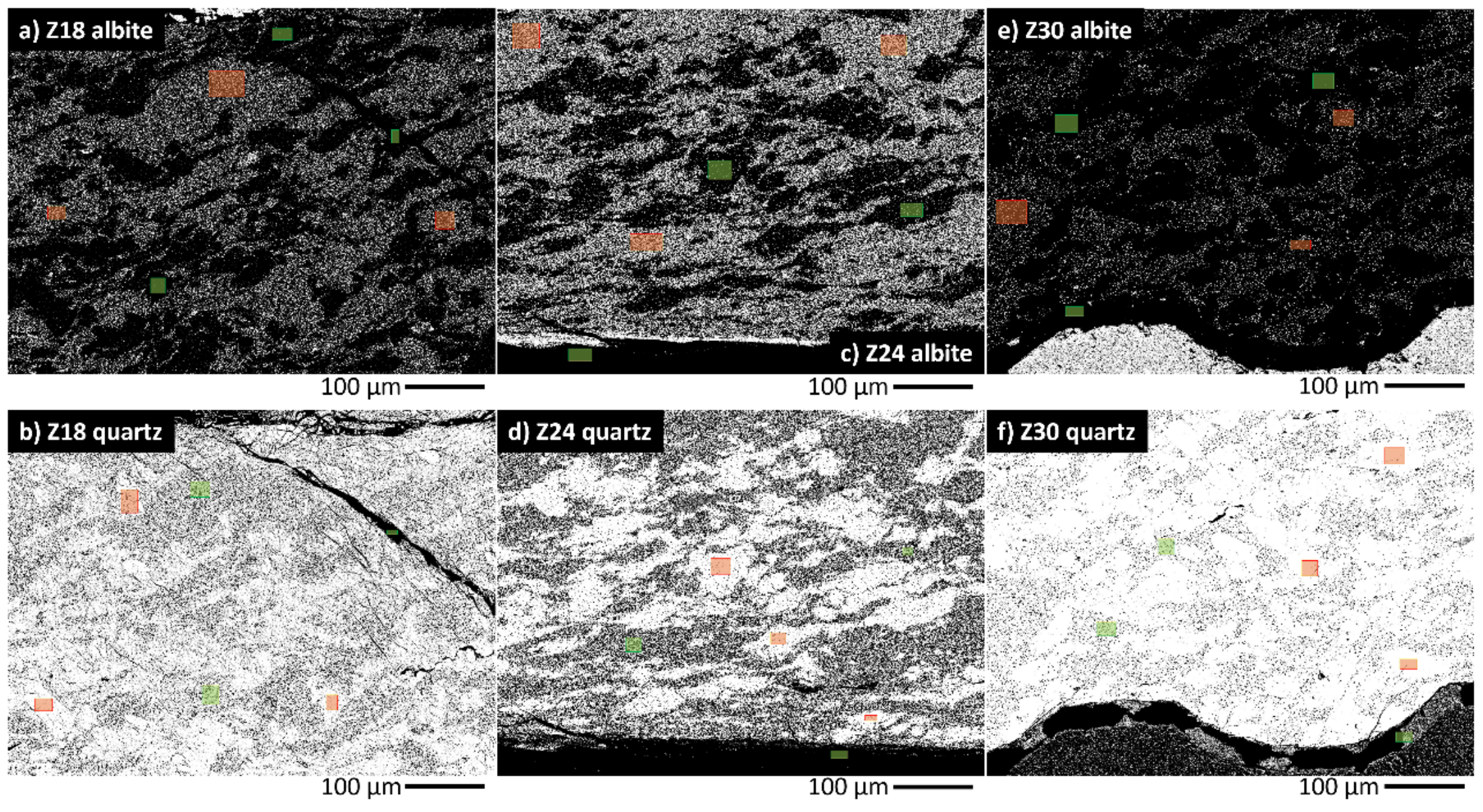

Figure S4. Phase plotting maps with the trained regions of the machine learning performed for the phase segmentation. Red rectangles denote the areas trained as Class 1, and green rectangles denote the areas trained as Class 2. (a) Albite map of sample Z18. (b) Quartz map of sample Z18. (c) Albite map of sample Z24. (d) Quartz map of sample Z24. (e) Albite map of sample Z30. (f) Quartz map of sample Z30. The phase to be analyzed (i.e., albite in Figures S4a, S4c, and S4e, and quartz in Figures S4b, S4d, and S4f) is shown in white.

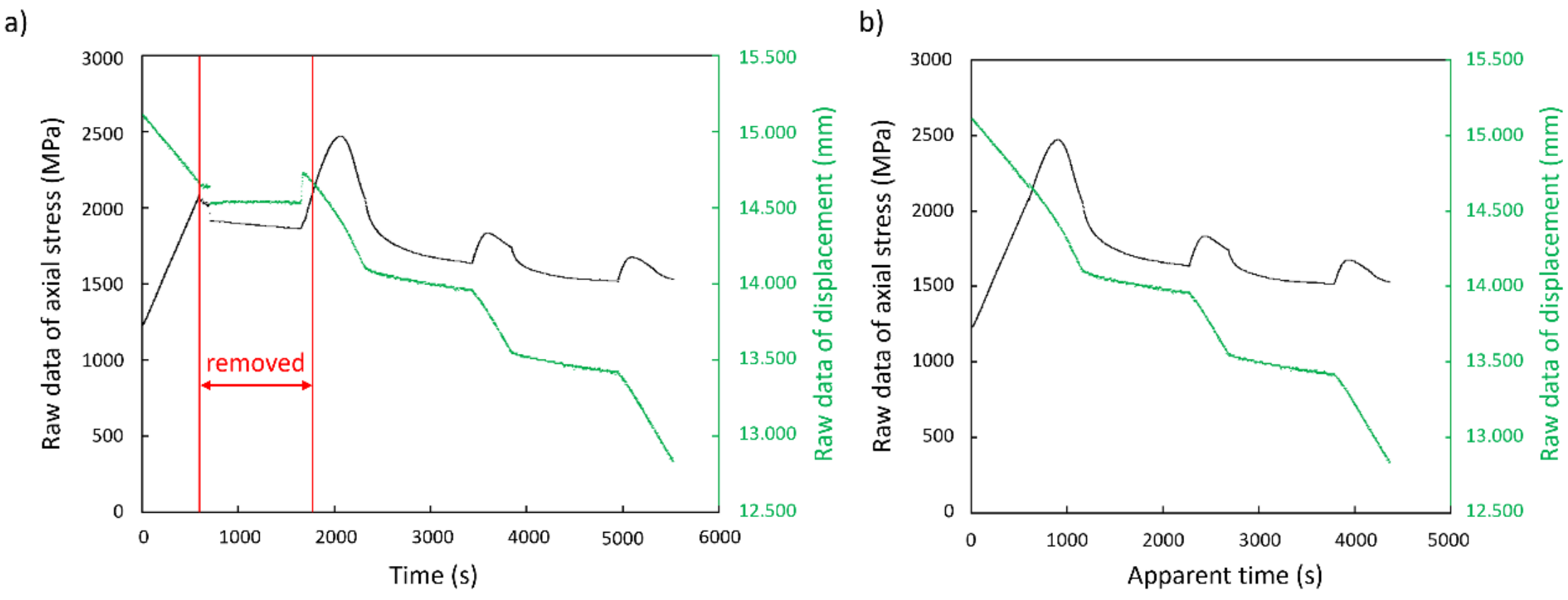

Figure S5. Raw data of experiment Z30. (a) Axial stress (black) and displacement (green) with time. The removed period of data are shown between the red lines. (b) Axial stress (black) and displacement (green) after removing the data of the period indicated in Figure

S5a, which were used for the calculations of shear stress and shear strain. Note that the horizontal axis shows the apparent time which differ from the actual time shown in Figure S5a, as it was generated by connecting the data before and after the removed part.

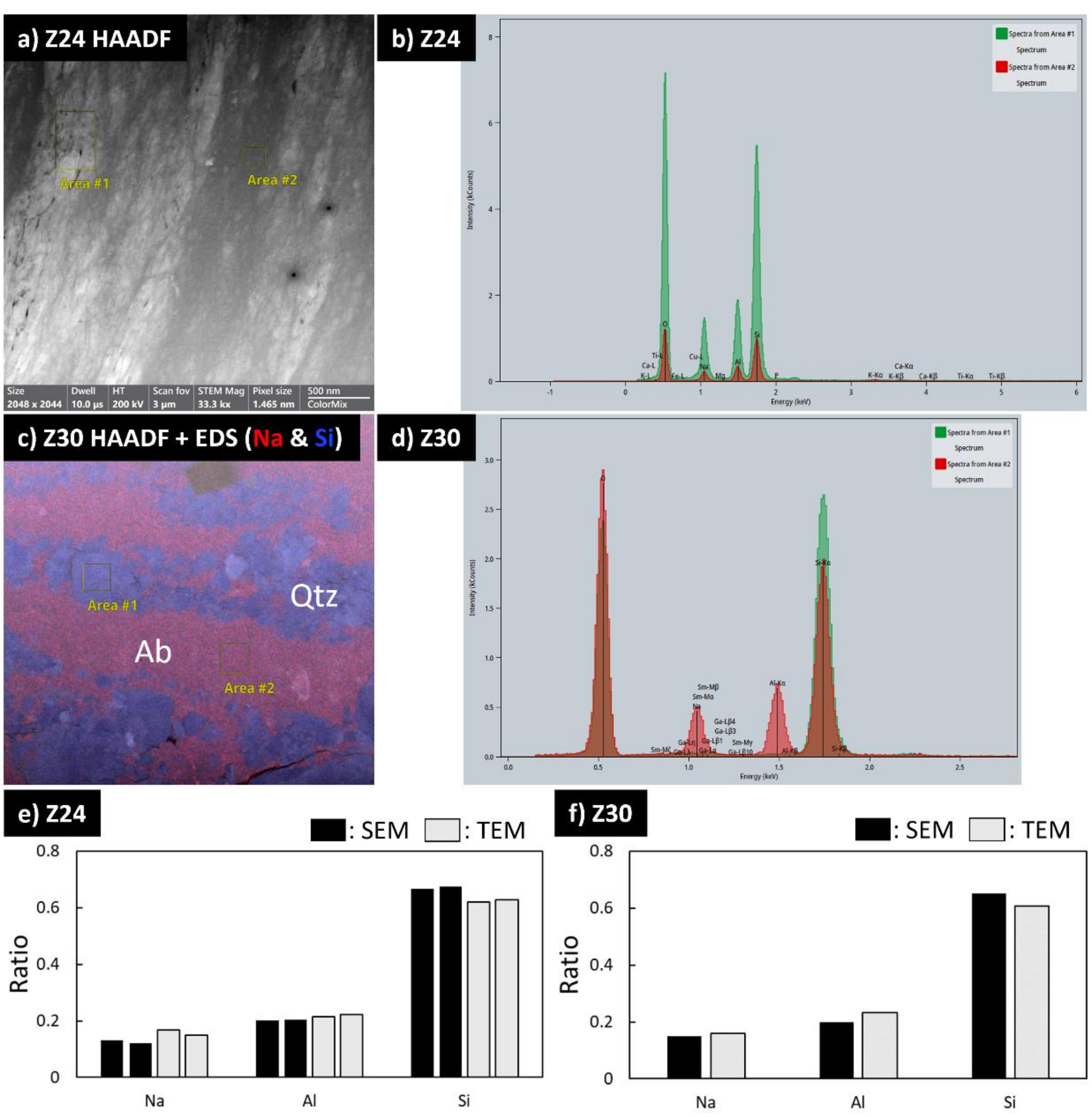

Figure S6. Comparisons of the compositions at the albite domains in samples Z24 and Z30. (a) HAADF image of sample Z24 with rectangles showing the areas where the EDS spectra in Figure S6b were taken. (b) EDS spectra in the albite domain in sample Z24. (c) HAADF image overlapped by element maps of Na (red) and Si (blue) in sample Z30. Qtz: quartz, Ab: albite. The rectangles show the areas where the EDS spectra in Figure S6d were taken. (d)

EDS spectra taken at the quartz domain (green) and the albite domain (red) in sample Z30. (e and f) Comparisons of the Na-Al-Si ratios calculated from the results of SEM and TEM analyses of sample Z24 (Figure S6e) and sample Z30 (Figure S6f). The ratios of SEM were calculated using the values of the approximate concentration of the elements, while those of TEM were calculated using the heights of the EDS spectra in Figures S6b and S6d. The black bars denote the results of SEM analyses, and the gray bars denote the results of the TEM analyses.

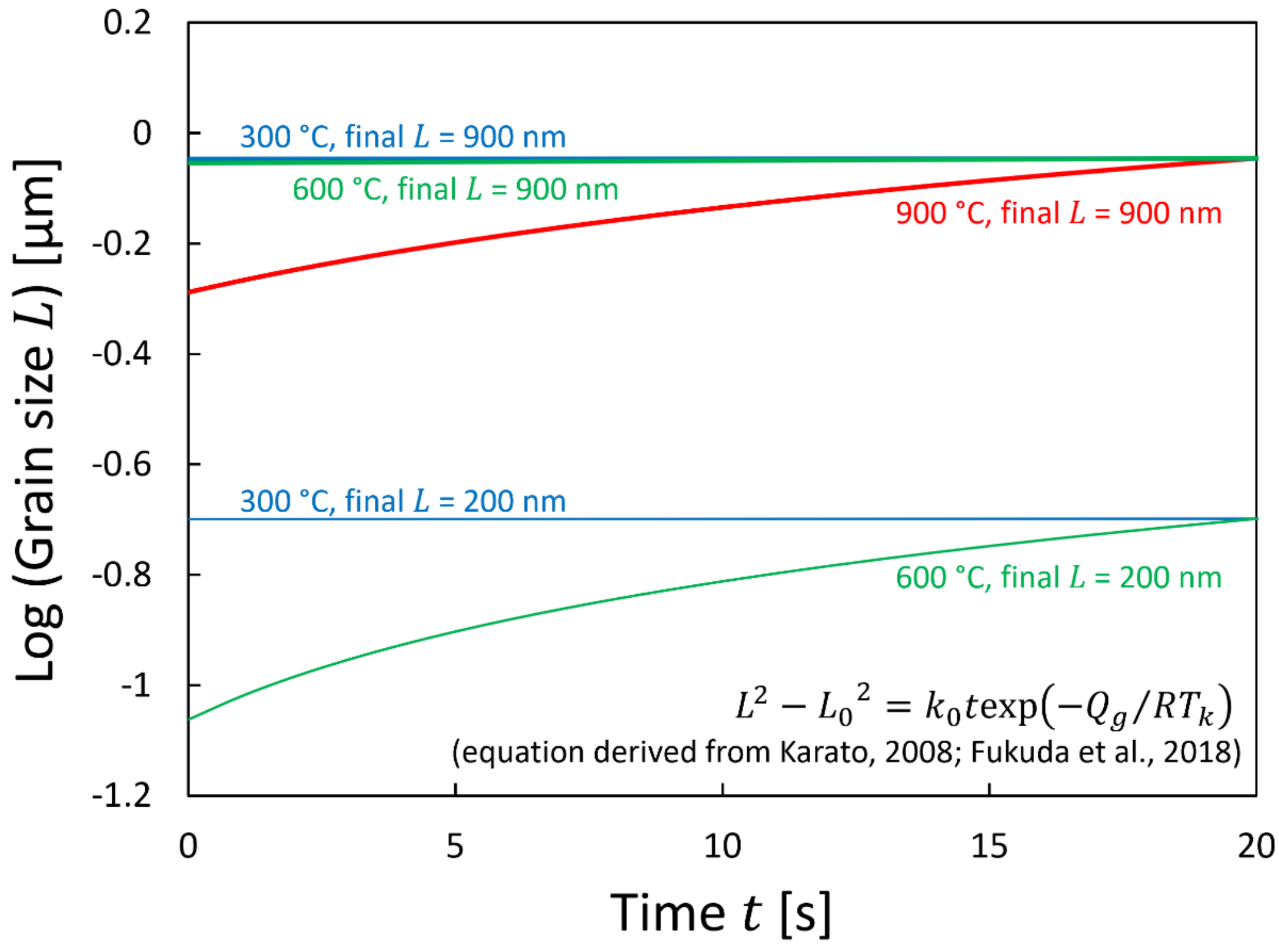

Fig. S7 Furukawa et al. (2025)

Figure S7. The semi-log plots of the relationships between the calculated quartz grain size $L$ after a certain time $t$ of growth during the quench of experiment Z30. We used the grain growth law of wet quartz (Karato, 2008, p.237 & p.241; Fukuda et al., 2018) as denoted in the lower right part of the figure, where $L_0$ is the initial grain size [µm], $k_0$ is a constant factor [µm$^2$/s], $Q_g$ is activation energy [kJ/mol], $R$ is the gas constant [kJ/(K·mol)], and $T_k$ is absolute temperature [K]. The thick lines denote the growth of the grains whose sizes are 900 nm at the end of the quench, while the thin lines denote that of the grains whose sizes

are 200 nm at the end of the quench. The growth at a temperature of 900 ° C is shown in red, that of 600 ° C is shown in green, and that of 300 ° C is shown in blue.